\makeatletter
\def\cl@chapter{}
\makeatother
\RequirePackage{fix-cm}
\documentclass[smallextended,numbook]{svjour3}

\usepackage[table]{xcolor}

\usepackage{svg}      \usepackage{tabulary} \usepackage{multirow} \usepackage{tabularx}
\usepackage{xparse}
\usepackage{tikz}

\usepackage{cite}
\usepackage{adjustbox}

\usepackage{booktabs}
\usepackage[T1]{fontenc}
\usepackage[figuresright]{rotating}
\usepackage[noindentafter]{titlesec}
\titlespacing{\paragraph}{0pt}{\smallskipamount}{4pt}
\titleformat{\paragraph}[runin]{\itshape}{}{}{> }[~---] \titleformat{\subsubsection}[block]{\normalfont}{\arabic{section}.\arabic{subsection}.\arabic{subsubsection}}{3pt}{} 
\usepackage{enumitem}
\usepackage{noindentafter}
\NoIndentAfterEnv{itemize}
\NoIndentAfterEnv{description}
\NoIndentAfterEnv{enumerate}
\setlist{
	leftmargin=12pt,
	itemsep=2pt plus 1pt,
	topsep=2pt plus 1pt,
parsep=2pt plus 1pt,
	listparindent=0pt,}
\setlist[description]{
	font=\normalfont\itshape
}
\makeatletter
\expandafter\patchcmd\csname end \endcsname{\if@ignore\@ignorefalse\ignorespaces\fi }{\if@ignore\@ignorefalse\ignorespaces\fi \csuse{@noindent@#1@hook}}{}{\PackageWarningNoLine{noindentafter}{Patching `\string\end' failed!\MessageBreak `\string\NoIndentAfter...' commands won't work}}
\makeatother

\NewDocumentCommand\name{+m}{{\small\scshape #1}\xspace}

\usepackage{xspace}
\usepackage{adjustbox}
\usepackage{cleveref}
\Crefname{section}{Sec.}{Sec.}
\Crefname{figure}{Figure}{Figures}
\usepackage{url}

\newcommand{\etal}{et al.}

\usepackage{framed}
\usepackage{xparse}
\newsavebox{\fminipagebox}
\NewDocumentEnvironment{hassanbox}{m O{\fboxsep}}
 {\par\kern#2\noindent\begin{lrbox}{\fminipagebox}
  \begin{minipage}{#1}\ignorespaces}
 {\end{minipage}\end{lrbox}\makebox[#1]{\kern\dimexpr-\fboxsep-\fboxrule\relax
    \fbox{\usebox{\fminipagebox}}\kern\dimexpr-\fboxsep-\fboxrule\relax
  }\par\kern#2
 }

\newcommand{\tbdWlabelledset}{the labelled app stores\xspace}

\newcommand{\AppAstores}{an app store\xspace}
\newcommand{\AppStores}{App Stores\xspace}
\newcommand{\AppStore}{App Store\xspace}
\newcommand{\kmeans}{\emph{K-means}\xspace}
\newcommand{\appStores}{app stores\xspace}
\newcommand{\appStore}{app store\xspace}

\newcommand{\googlePlay}{\name{Google Play}}
\newcommand{\appleAppStore}{Apple's \name{App Store}}
\newcommand{\android}{\emph{Android}\xspace}

\definecolor{progresscolor}{RGB}{184,55,132}

\definecolor{wzcolor}{RGB}{255,140,0}
\newcommand{\wz}[1]{{\color{wzcolor}Cosmos: #1}\xspace}
\usepackage[normalem]{ulem}

\definecolor{mikecolor}{RGB}{255,140,0}

\newcommand{\step}[1]{Step~\textcircled{\footnotesize #1}}
\newcommand{\stepn}[1]{\textcircled{\footnotesize #1}}

\newcommand{\domain}[1]{\emph{\url{#1}}}

\newcommand{\tbdSrqone}{\textbf{RQ1}: \emph{What fundamental \tbdWdims describe the space of app stores?}}
\newcommand{\tbdSrqtwo}{\textbf{RQ2}: \emph{Are there groups of stores that share similar \tbdWdims?}}

\newcommand{\tbdWdim}{feature\xspace}
\newcommand{\tbdWdims}{features\xspace}

\newcommand{\tbdWDims}{Features\xspace}

\newcommand{\tbdVbestK}{8\xspace}

\newcommand{\tbdVlabelledStores}{53\xspace}

\begin{document}

\title{What is an App Store? \\ The Software Engineering Perspective}

\author{Wenhan Zhu \and
	Sebastian Proksch \and
	Daniel M. German \and
	Michael W. Godfrey \and
	Li Li \and
	Shane McIntosh
}

\institute{Wenhan Zhu \and Michael W. Godfrey \and Shane McIntosh
          \at David R. Cheriton School of Computer Science, University of waterloo,
          Waterloo, Canada \\\email{\{w65zhu, migod, shane.mcintosh\}@uwaterloo.ca}
          \and
          Sebastian Proksch
          \at Delft University of Technology,
          Delft, Netherlands \\\email{s.proksch@tudelft.nl}
          \and Daniel M. German
          \at Department of Computer Science, University of Victoria,
          Victoria, Canada \\\email{dmg@uvic.ca}
          \and Li Li
          \at Faculty of Information Technology, Monash University,
          Melbourne, Australia \\\email{li.li@monash.edu}
}

\maketitle

\begin{abstract}

    ``App stores'' are online software stores where end users may browse,
    purchase, download, and install software applications. By far, the best
    known app stores are associated with mobile platforms, such as
    \googlePlay for \android and \appleAppStore for iOS.  The ubiquity of
    smartphones has led to mobile app stores becoming a touchstone
    experience of modern living. App stores have been the subject of many
    empirical studies.
    However, most of this research has concentrated on properties of the
    apps rather than the stores themselves.  Today there is a rich
    diversity of app stores, and these stores have largely been overlooked
    by researchers: app stores exist on many distinctive platforms, are
    aimed at different classes of users, and have different end-goals
    beyond simply selling a standalone app to a smartphone user.

    The goal of this paper is to survey and characterize the broader
    dimensionality of app stores, and to explore how and why they 
    influence software development practices, such as system design
    and release management.  We begin by collecting a set of app store
    examples from web search queries.  By analyzing and curating the
    results, we derive a set of \tbdWdims common to app stores.  We then
    build a dimensional model of app stores based on these \tbdWdims, and
    we fit each app store from our web search result set into this model.
    Next, we performed unsupervised clustering to the app stores to find
    their natural groupings. Our results suggest that app stores have
    become an essential stakeholder in modern software development.  They
    control the distribution channel to end users and ensure that the
    applications are of suitable quality; in turn, this leads to developers
    adhering to various store guidelines when creating their applications.
    However, we found the app stores’ operational model could vary widely
    between stores, and this variability could in turn affect the
    generalizability of existing understanding of app stores.

\end{abstract}


\section{Introduction}

The widespread proliferation of smartphones and other mobile devices in recent years has in turn produced an immense demand for applications that run on these platforms.
In response, online ``app stores'' such as \googlePlay and \appleAppStore have emerged to facilitate the discovery, purchasing, installation, and management of apps by users on their mobile devices.
The success of mobile app stores has enabled a new and more direct relationship between app creators and users.
The app store serves as a conduit between software creators (often, developers) and their users, with some mediation provided by the app store.
The app store provides a ``one-stop shopping'' experience for users, who can compare competing products and read reviews of other users.
The app store might also acts as a quality gatekeeper for the platform, providing varying levels of guarantees about the apps, such as easy installation and removal, expected functionality, and malware protection.
To the software creator, the app store provides a centralized marketplace for their app, where potential users can find, purchase, and acquire the app easily; the app store also relieves the developer from basic support problems related to distribution and installation, since apps must be shown to install easily during the required approval process.
Indeed, one of the key side effects of mobile app stores is that it has forced software developers to streamline their release management practices and ensure hassle-free deployment at the user's end.

The success of mobile app stores has also led to the establishment of a plethora of other kinds of app store, often for non-mobile platforms, serving diverse kinds of user communities, offering different kinds of services, and using a variety of monetization strategies.
Many technical platforms now operate in a \emph{store-centric} way: essential services and functionality are provided by the platform while access to extensions/add-ons is offered only through interaction with the app store.
For instance, \googlePlay, the app store, operates on top of the technical
platform \android, which provides the runtime environment for the applications.
When new technical platforms are introduced, an app store is often expected to serve as a means to host and deliver products to its users~\cite{dixon2010home}.
Example technical platforms that use app store-like approaches include \name{Steam}~\cite{store.steam}, \name{GitHub Marketplace}~\cite{store.github}, the \name{Chrome Web Store}~\cite{store.chrome}, \name{WordPress}~\cite{store.wordpress}, \name{AutoDesk}~\cite{store.autodesk}, \name{DockerHub}~\cite{store.dockerhub}, \name{Amazon Web Services} (\name{AWS})~\cite{store.aws}, \name{Homebrew}~\cite{store.homebrew}, or \name{Ubuntu Packages}~\cite{store.ubuntu}.

For platforms that operate in this way, the app store is an essential part of the platform's design.
For example, consider source code editors, such as \name{VSCode} and \name{IntelliJ}.  
The tool itself --- which we consider to be a technical platform in this context --- offers the essential functionality of a modern source code editor; however, many additional services are available through the associated app store that are not included by default.  
Thus, extensions that allow for language-specific syntax highlighting or version control integration must be added manually by the user through interaction with the tool's app store.
We conjecture that the app store has fundamentally changed how some classes of software systems are designed, from the overall ecosystem architecture of the technical platform to the way in which add-ons are engineered to fit within its instances.  

In this work, we will explore the general space of app stores, and also consider how app store-centric design can affect software development practices.
Previous research involving app stores has focused mainly on mobile app stores, often concentrating on properties of the apps rather than properties of the stores.  
For example, Harman et al.\ performed one of the first major studies of app
stores in 2012, focusing on the \name{BlackBerry App World}~\cite{harman2012app}.
However, concentrating the investigative scope so narrowly may lead to claims that do not generalize well across the space of all app stores.  
For example, Lin et al.\ found that reviews of games that appeared in mobile app stores differed significantly from the reviews of the same game that appeared within the \name{Steam} platform's own app store~\cite{lin2019empirical}.
In our work, we aim to take a more holistic approach to studying app stores by considering both mobile and non-mobile variants.
In so doing, we hope to create a more general model of app stores that fits this broader space.

To achieve a holistic view, we start from the definition of an app store.
A precise definition of the term ``app store'' has been omitted in much of the previous research in this area.
Currently, \googlePlay and \appleAppStore dominate the market and are the main targets of research on app stores; in the past, the \name{BlackBerry App World} and Microsoft's \name{Windows Phone Store} were also important players, but these stores are now defunct.\footnote{The \name{Windows Phone Store} was absorbed into the broader \name{Windows Store} in 2015.}
Wikipedia recognizes \name{\footnotesize Electronic AppWrapper}~\cite{wiki.appwrapper} as the first true platform-specific electronic marketplace for software applications, but the term became popular when Apple introduced its \name{App Store} along with the iPhone 3G in 2008.
Since then, the term has largely come to refer to any centralized store for mobile applications.
We present our own working definition of the term ``app store'' in \Cref{sec:app-store-def}.

The goal of this work is to survey and characterize the broader dimensionality of app stores, and also to explore how and why they may feed back into software development practices, such as release management.
As a step toward this goal, we focus on two research questions (RQs) that aim to explore the space of app stores:

\begin{description}
  \item{\tbdSrqone} 
\end{description}
To understand app stores, we first need a way to describe them.
  It would be especially useful if this description framework would highlight the similarities and differences of app stores.
We start by collecting a set of app store examples, and then extract from them a set of \tbdWdims that illustrate important differences between them.
  We then expand this list of app stores with search queries to derive a larger set of example stores.
  We explicitly seek generalized web queries to broaden our search space beyond the common two major mobile app stores of \emph{Apple} and \emph{Google}.
  By combining the web queries and the initial set of app stores, we selected a representative set of app stores and extracted their \tbdWdims.
In the end, we first surveyed app stores and derived a \tbdWdim-based model to describe them; we then expanded the set of app store through web queries; and finally, we extracted \tbdWdims based on the model for a representative set of app stores.

 \begin{description}
\item{\tbdSrqtwo}
\end{description}
Despite the ability to describe individual stores, it is also important to understand the relationships between different stores.
  Having a understanding of the natural groupings can help us gain insights into the generalizability of results gathered for app stores.
We perform a \kmeans~\cite{macqueen1967some} clustering based on the extracted \tbdWdims of the expanded set of app stores collected previously.
  The optimal $k$ value is determined by the Silhouette method~\cite{rousseeuw1987silhouettes}.
The clustering results suggest that there are 8 groups in the expanded set of app stores.
  The differences can be observed in the type of application offered, standalone or extension, and/or type of operation, business or community-oriented.

In this study, we make several contributions towards a better understanding of the app store ecosystem.
\begin{itemize}[noitemsep]
    \item  We identified a set of descriptive features that can be used to characterize app stores.
    \item We identified a set of 291 app stores and mapped \tbdVlabelledStores of them into the feature space.
    \item We identified \tbdVbestK coherent groups of app stores based on the similarity of features.
    \item We discuss our insights on how the features and the diversity of app stores can impact software engineering.
\end{itemize}
Overall, our study contributes towards a holistic view of app stores within software engineering, which can form the basis for subsequent study of app stores in general.

\section{Background and Related Work}
\label{sec:background}

\subsection{Early App Store Research}

To date, research in this area has concentrated on a narrow set of app stores that primarily involves mobile platforms.
Harman~\etal{}~\cite{harman2012app}~proposed app stores as a valid kind of software repository worthy of formal study within the broader research area of mining software repositories; while their work was not specific to mobile app stores, they used \name{BlackBerry App World} as their canonical example.  
Ruiz~\etal{}~\cite{ruiz2012understanding} studied the topic of reuse within app stores, focusing their work on \name{Android Marketplace}.\footnote{\name{Android Marketplace} has since been re-branded as \googlePlay.}
In both cases, these early works did not provide a formal definition of ``app store'', and tacitly used only app stores for mobile platforms in their studies.

In their 2016 survey on app store research, Martin~\etal{}~\cite{martin2016survey} observed that studies have often focused on only a few specific app stores, and have ignored comparisons between app stores.
In a recent literature survey, D{\k{a}}browski~\etal{}~\cite{dkabrowski2022analysing} found the median number of app stores studied to be 1, with the maximum being 3.
We also note that results from one app store study may not generalize to another store since the two stores may differ in significant ways; for example, if a store does not allow users to provide their own reviews of the apps within the store, app creators will have to rely on other means to gain popularity and trust from users, such as promotion outside of the app store.
The same trend can be observed in more specific app store topics such as app reviews; for example, Lin~\etal{}~\cite{lin2019empirical}~found that reviews of games within the \name{Steam} app store can be dramatically different from reviews of the same game in mobile app stores.

Existing work has yet to explore the full diversity of app stores, concentrating on \googlePlay and \appleAppStore, and largely ignoring those such as \name{Steam}, \name{AWS}, and \name{GitHub Marketplace} that are not specific to mobile platforms.
With the heterogeneity of app stores and their typical uses, we believe that the research in this area can be strengthened by expanding the breadth to encompass a more diverse perspective on app stores; in turn, this breadth can help to validate the generalizability of the study findings.

\subsection{App Stores in Recent Software Engineering Research}

To better understand the involvement of app stores in recent research, we reviewed relevant recent papers from the two flagship software engineering research conferences: the \emph{ACM/IEEE International Conference on Software Engineering} (``ICSE'')
and the \emph{ACM SIGSOFT International Symposium on the Foundations of Software Engineering} (``FSE'')
We used \emph{Google Scholar} to find papers containing the keyword ``app store'' between January 2020 and April 2022 for the two conferences.
We found a total of 34 such papers (listed in \Cref{tab:app_store_paper}).
After reading through all of them, we found that each paper fit into one of two broad categories:  \textit{mining software applications} (20/34) and \textit{mining app store artifacts} (14/34).
We note that our efforts do not constitute a comprehensive literature survey; instead, our goal was to gain an overview of how app stores are involved in recent research, and why app stores matter in their context.

\begin{table*}[ht]
  \centering
  \footnotesize
  \caption{Recent papers on app stores}
  \newcommand{\wzIndentWidth}{2mm}
  \label{tab:app_store_paper}
  \begin{adjustbox}{max width=\textwidth}
  \begin{tabular}{@{}lp{105mm}p{40mm}@{}}
    \toprule
    Loc & Paper & Store \\
    \midrule
    \multicolumn{3}{@{}l}{\bf Mining software applications} \\
    \hspace{\wzIndentWidth}ICSE~'21 & {Atvhunter: Reliable version detection of third-party libraries for vulnerability identification in android applications}~\cite{zhan2021atvhunter} & \googlePlay \\
    \hspace{\wzIndentWidth}ICSE~'20 & {How does misconfiguration of analytic services compromise mobile privacy?}~\cite{zhang2020does} & \googlePlay \\
    \hspace{\wzIndentWidth}FSE~'21 & {Algebraic-datatype taint tracking, with applications to understanding Android identifier leaks}~\cite{rahaman2021algebraic} & \googlePlay \\
    \hspace{\wzIndentWidth}FSE~'20 & {Code recommendation for exception handling}~\cite{nguyen2020code} & \googlePlay \\
    \hspace{\wzIndentWidth}FSE~'20 & {Static asynchronous component misuse detection for Android applications}~\cite{pan2020static} & F-Droid, \googlePlay, Wandoujia App Store \\
    \hspace{\wzIndentWidth}ICSE~'21 & {Sustainable Solving: Reducing The Memory Footprint of IFDS-Based Data Flow Analyses Using Intelligent Garbage Collection}~\cite{arzt2021sustainable} & \googlePlay \\
    \hspace{\wzIndentWidth}ICSE~'22 & {DescribeCtx: Context-Aware Description Synthesis for Sensitive Behaviors in Mobile Apps}~\cite{yao2022describectx} & \googlePlay \\
    \hspace{\wzIndentWidth}ICSE~'20 & {Time-travel testing of android apps}~\cite{dong2020time} & \googlePlay \\
    \hspace{\wzIndentWidth}ICSE~'20 & {An empirical assessment of security risks of global android banking apps}~\cite{chen2020empirical} & \googlePlay, APKMonk, and others \\
    \hspace{\wzIndentWidth}ICSE~'21 & {Too Quiet in the Library: An Empirical Study of Security Updates in Android Apps’ Native Code}~\cite{almanee2021too} & \googlePlay \\
    \hspace{\wzIndentWidth}ICSE~'20 & {Accessibility issues in android apps: state of affairs, sentiments, and ways forward}~\cite{alshayban2020accessibility} & \googlePlay \\
    \hspace{\wzIndentWidth}ICSE~'21 & {Don’t do that! hunting down visual design smells in complex uis against design guidelines}~\cite{yang2021don} & Android \\
    \hspace{\wzIndentWidth}ICSE~'21 & {Identifying and characterizing silently-evolved methods in the android API}~\cite{liu2021identifying} & \googlePlay \\
    \hspace{\wzIndentWidth}ICSE~'21 & {Layout and image recognition driving cross-platform automated mobile testing}~\cite{yu2021layout} & \appleAppStore, \googlePlay \\
    \hspace{\wzIndentWidth}FSE~'21 & {An empirical study of GUI widget detection for industrial mobile games}~\cite{ye2021empirical} & Android Games \\
    \hspace{\wzIndentWidth}ICSE~'21 & {Fine with “1234”? An Analysis of SMS One-Time Password Randomness in Android Apps}~\cite{ma2021fine} & \googlePlay, Tencent Myapp \\
    \hspace{\wzIndentWidth}ICSE~'21 & {IMGDroid: Detecting Image Loading Defects in Android Applications}~\cite{song2021imgdroid} & Android \\
    \hspace{\wzIndentWidth}ICSE~'21 & {GUIGAN: Learning to Generate GUI Designs Using Generative Adversarial Networks}~\cite{zhao2021guigan} & Android \\
    \hspace{\wzIndentWidth}ICSE~'20 & {Unblind your apps: Predicting natural-language labels for mobile gui components by deep learning}~\cite{chen2020unblind} & \googlePlay \\
    \hspace{\wzIndentWidth}FSE~'21 & {Frontmatter: mining Android user interfaces at scale}~\cite{kuznetsov2021frontmatter} & \googlePlay \\
    \multicolumn{3}{@{}l}{\bf Mining app store non-technical attributes} \\
    \hspace{\wzIndentWidth}ICSE~'20 & {Schr{\"o}dinger's security: Opening the box on app developers' security rationale}~\cite{van2020schrodinger} & \appleAppStore, \googlePlay \\
    \hspace{\wzIndentWidth}ICSE~'20 & {Scalable statistical root cause analysis on app telemetry}~\cite{murali2021scalable} & Facebook App \\
    \hspace{\wzIndentWidth}ICSE~'21 & {An empirical assessment of global COVID-19 contact tracing applications}~\cite{sun2021empirical} & Android \\
    \hspace{\wzIndentWidth}ICSE~'21 & {We’ll Fix It in Post: What Do Bug Fixes in Video Game Update Notes Tell Us?}~\cite{truelove2021we} & Steam \\
    \hspace{\wzIndentWidth}ICSE~'21 & {Automatically matching bug reports with related app reviews}~\cite{haering2021automatically} & \googlePlay \\
    \hspace{\wzIndentWidth}ICSE~'21 & {Prioritize crowdsourced test reports via deep screenshot understanding}~\cite{yu2021prioritize} & Android \\
    \hspace{\wzIndentWidth}ICSE~'21 & {A first look at human values-violation in app reviews}~\cite{obie2021first} & \googlePlay \\
    \hspace{\wzIndentWidth}ICSE~'21 & {Does culture matter? impact of individualism and uncertainty avoidance on app reviews}~\cite{fischer2021does} & \appleAppStore \\
    \hspace{\wzIndentWidth}ICSE~'21 & {COVID-19 vs social media apps: does privacy really matter?}~\cite{haggag2021covid} & \googlePlay, \appleAppStore \\
    \hspace{\wzIndentWidth}ICSE~'20 & {Society-oriented applications development: Investigating users’ values from bangladeshi agriculture mobile applications}~\cite{shams2020society} & \googlePlay \\
    \hspace{\wzIndentWidth}FSE~'21 & {Checking conformance of applications against GUI policies}~\cite{zhang2021checking} & Android \\
    \hspace{\wzIndentWidth}ICSE~'21 & {Identifying key features from app user reviews}~\cite{wu2021identifying} & \appleAppStore \\
    \hspace{\wzIndentWidth}ICSE~'21 & {Champ: Characterizing undesired app behaviors from user comments based on market policies}~\cite{hu2021champ} & \googlePlay, Chinese android app stores \\
    \hspace{\wzIndentWidth}ICSE~'20 & {Caspar: extracting and synthesizing user stories of problems from app reviews}~\cite{guo2020caspar} & \appleAppStore \\
    \bottomrule
  \end{tabular}
\end{adjustbox}
\end{table*}

\paragraph{Mining software applications} App stores have been extensively used as a mining source of software applications. 
In these papers, the major focus is often on another subject and app stores provide a source where they can collect applications for either a data source or verification dataset.
For example, Zhan~\etal{}~\cite{zhan2021atvhunter} proposed an approach to
detect software vulnerabilities in third-party libraries of \android applications.
They leveraged the app store to collect a dataset to verify the effectiveness of their approach.
In these studies, the app store is both a convenient and practical source of data collection.
However, the involvement of app stores may not be necessary since the purpose is to gather a dataset of application.
In Yang~\etal{}'s work~\cite{yang2021don}, they leveraged \android applications from an existing dataset without the need to collect from an app store.
We argue that the importance of app stores in these types of studies is the selection criteria performed by the researchers to collect applications from app stores.
These features can include star ratings, total downloads, app category.

\paragraph{Mining app store artifacts} In these studies, researchers focused on unique software artifacts that come from the operation of the app stores. 
App stores have a much heavier involvement in these studies compared to the previous group.
App reviews is the major software artifact the researchers focused on, where they leverage the data to identify features of applications~\cite{wu2021identifying}, locating bug reports~\cite{haering2021automatically}, and detect undesired app behaviors~\cite{hu2021champ}.
One interesting research practice we observed is where van der Linden~\etal{}~\cite{van2020schrodinger} leveraged the developer contact information shared on app stores to send out surveys related to security practices.

\subsection{Store-Focused Research}

As stated above, we found that most recent research involving app stores focuses on the applications they offer rather than on studying the app stores themselves; in particular, most research in the domain focuses on the development of mobile applications.
Meanwhile, a few papers have specifically considered app stores and their effects on software engineering,
but again these works focus heavily on mobile app stores.

In a recent paper, Al-Subaihin~\etal{}~\cite{al-subaihin2021app} interviewed developers about how app stores affect their software engineering tasks.
They found that developers often leverage the review section from similar applications to help with understanding the expected user experience and anticipated features.
App stores also provides a playground for releasing beta version of apps to receive feedback from users.
The built-in communication channels also play a large role in informing development.
The interviews suggest that developers pay attention to viewing user requests in app store via channels such as reviews and forums.
The approval period of app stores affects how developers plan their release.
App stores introduce non-technical challenges in the development process.
Given the app store model of release, app store-specific metrics, such as total number of downloads, are considered highly important to developers.

Running an app store presents both technical and non-technical challenges to the store owner.
Technical challenges include verifying that each app will install correctly, while non-technical challenges include ensuring that the promotional information in the app's product page adheres to store guidelines.
Wang~\etal{}~\cite{wang2018beyond} investigated several \android app stores in China and compared them to \googlePlay.
Their study showed that these stores were much less diligent in screening the apps they offered, with a significantly higher presence of fake, cloned, and malicious apps than \googlePlay.

Jansen and Bloemendal surveyed the landscape of app stores from the perspective of the business domain~\cite{jansen2013defining}.
They selected 6 app stores (5 mobile stores and 1 \emph{Windows} store) at the time of publication (2013), and investigated each store
manually to find features (those actors can interact with) and policies (rules, regulations and governing processes that
limit the functional reach of the features) from each app store. The actors they define are the same as the three major
stakeholders of the app store model (i.e., the store owner, users, and developers). Our study further contributes
to the understanding of app-stores. First, we studied a significantly larger set of app-stores: our methodology was focused towards the identification of as
many different types of stores as possible. In total we studied \tbdVlabelledStores stores in various domains (such as
mobile, embedded systems, computer games, application add-ons, open source distributions and packaging systems, etc.).
Second, Jansen and Bloemendal studied app-stores from the perspective of a software business; for example, in their work they would consider features and policies on whether users are able to generate affiliate links to earn revenue through sharing applications.
In contrast, our work
focuses on app stores in the perspective of their role in the software engineering process.

In our study, we approach app stores from a broad landscape not limiting to mobile app stores.
We focus on the similarity of features offered between stores to understand their natural groupings and discuss the challenges in the diversity of app stores.

\subsection{Working Definition of an App Store}
\label{sec:app-store-def}

Previous researchers have often taken a casual approach to defining the term ``app store'', when a definition has been provided at all.
For example, in their survey paper, Martin et al.~ define an app store as ``A collection of apps that provides, for each app, at least one non-technical attribute'', with an app defined as ``An item of software that anyone with a suitable platform can install without the need for technical expertise''~\cite{martin2016survey}.
However, we feel that this definition is too generous.
For example, consider a static website called \name{Pat's Apps} that lists of a few of someone's (Pat's) favourite applications together with their personalized ratings and reviews; superficially, this would satisfy Martin et al.'s requirements as it is a collection of apps together with Pat's own reviews (which are non-technical attributes).
We feel that this kind of ``store'' is outside our scope of study for several reasons: Pat's software collection is not comprehensive, it is unlikely that Pat provides any technical guarantees about quality of the apps, and a passive list of apps on a web page does not constitute an automated ``store''.

\begin{figure}
\includegraphics[width=\columnwidth]{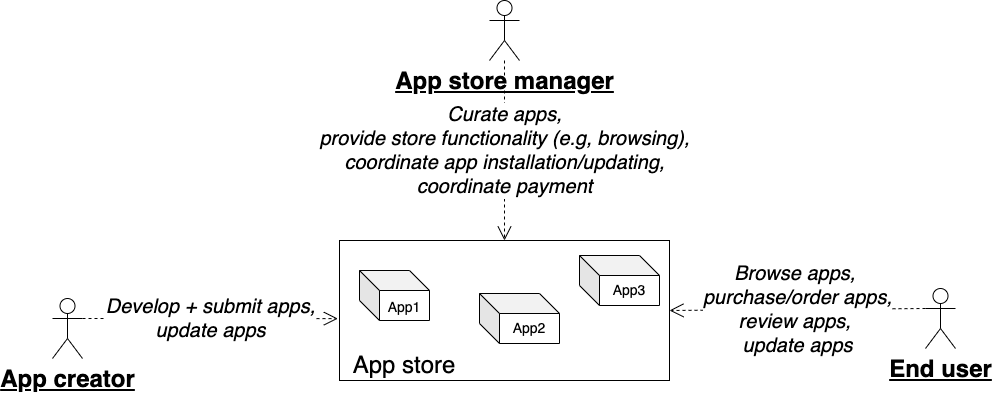}
    \caption{Three major stakeholders of most app stores}
\label{fig:store_stakeholder}
\end{figure}

Jansen and Bloemendal \cite{jansen2013defining} define app-store as ``An online curated marketplace that allows developers to sell and distribute their products to actors within one or more multi-sided software platform ecosystems.'' 
We note that this definition ignores that app-stores are expected to provide infrastructure for the deployment, installation, and maintenance of the apps, which impacts the software development process.
Their model also ignores marketplaces that do not have payment mechanisms, such as the Google Chrome Extensions store and the various open source apps stores, where all of the software products may be free to download and install.

In our work, we seek to define an idea of app store beyond the well-known mobile ones and with an emphasis on how their existence may affect the software development cycle.
Because we are focused on exploring the notion of what app stores are, we formulate a working definition of the term; we did so to provide clear inclusion/exclusion criteria for the candidate app stores that we discover in \Cref{sec:methodology}.

Our working definition was influenced by considering the three major stakeholders of the app store model: the \emph{app creators} who create and submit applications to the store; the \emph{app stores} themselves, and the organizations behind their operation who curate the app collection and coordinate both the store and installation mechanisms; and the \emph{end users} who browse, download, review, and update their applications through the app store (see \Cref{fig:store_stakeholder}).

We thus arrived at the following \textbf{working definition for \emph{app store}} as an online distribution mechanism that:
\begin{enumerate}[noitemsep]
\item offers access to a comprehensive collection of software or software-based services (henceforth, ``apps'') that augment an existing technical infrastructure (i.e., the runtime environment),
\item is curated, i.e., provides some level of guarantees about the apps, such as ensuring basic functionality and freedom from malware, and
\item provides an end-to-end automated ``store'' experience for end users, where
    \begin{enumerate}[noitemsep]
      \item the user can acquire the app directly through the store,
      \item users trigger store events, such as browsing, ordering, selecting options, arranging payment, etc., and
      \item the installation process is coordinated automatically between the store and the user's own instance of the technical platform.
    \end{enumerate}
\end{enumerate}
We can see that using this working definition, our \name{Pat's Apps} example fails to meet all three of our main criteria. 

We note that our working definition above evolved during our investigations; it represents our final group consensus on what is or is not an app store for the purposes of doing the subsequent exploratory study.
The steps by which the representation is finalized are discussed in \Cref{sec:methodology-stage-2}.
For example, our working definition implicitly includes \emph{package managers} such as the Debian-Linux \name{apt} tool and Javascript's \name{NPM} tool.
It is true that package managers are typically non-commercial, and so are ``stores'' only in a loose sense of the term; furthermore, they usually lack a mechanism for easy user browsing of apps and do not provide a facility for user reviews.
However, at the same time, they are a good fit conceptually: they tend to be comprehensive, curated, and offer an automated user experience for selection and installation.
Furthermore, some package managers serve as the backend to a more traditional store-like experience; for example, the \name{Ubuntu Software Center} builds on a tool \emph{aptitude}, which interacts with software repositories, to provide a user experience similar to that of \googlePlay.

\section{Research Methodology}
\label{sec:methodology}

\begin{figure}
  \centering
  \includegraphics[height=0.954\textheight]{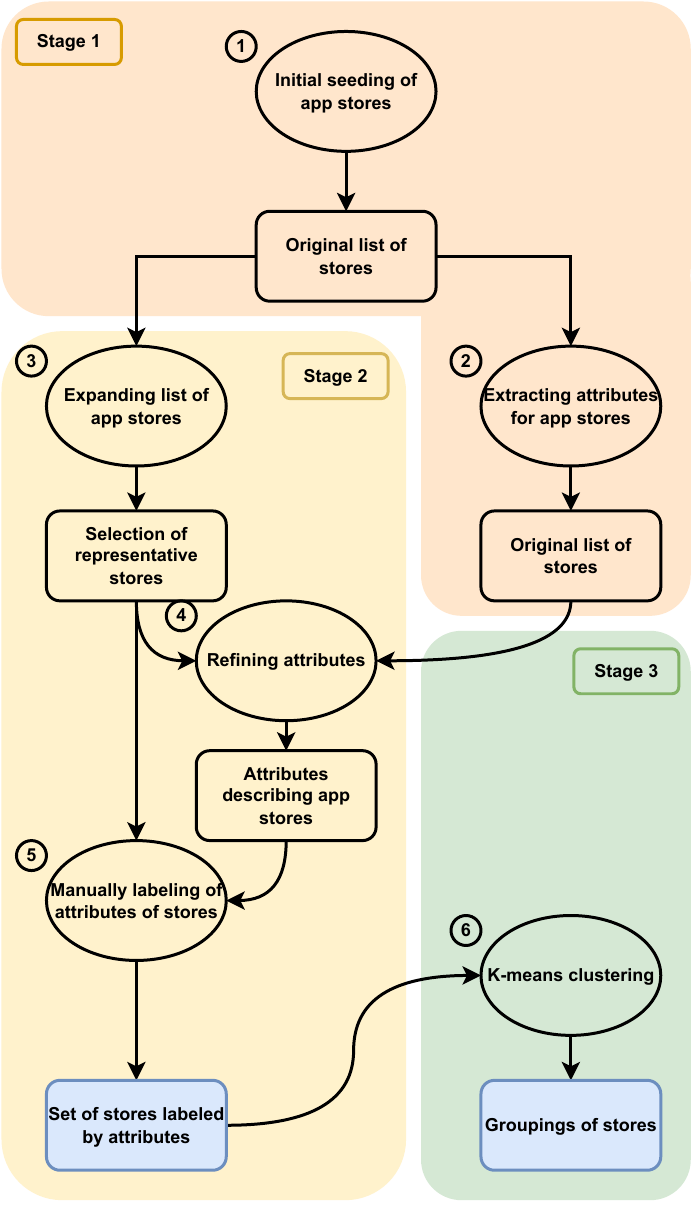}
  \caption{Methodology overview: There are three main \emph{stages}, further broken down into six \emph{steps}.}
\label{fig:method}
\end{figure}

To investigate the research questions, we designed a three-stage methodology that is illustrated in \Cref{fig:method}. The goal of the first two stages is to answer RQ1, while the third stage addresses RQ2.

In the first stage (\step{1} and \stepn{2}) we identified our initial list of \tbdWdims using a small set of well-known app stores (\appleAppStore, \googlePlay, \name{Steam} etc.)
In the second stage (Steps \stepn{3}, \stepn{4}, and \stepn{5}) we methodically expanded our list to a conceptually wider ranging set of \tbdVlabelledStores{} app stores.
We then described these stores using the \tbdWdims identified in the first stage.
A major goal of this stage was to evaluate whether the set of available \tbdWdims was sufficient to describe the characteristics of all these stores.
This set of \tbdWdims forms the answer to RQ1.

In the third stage (\step{6}), we took advantage of the labelling of the \tbdVlabelledStores{} stores.
We used \kmeans clustering analysis to identify groups of stores that shared similar \tbdWdims.
These groupings form the answer to RQ2.

We now describe our methodology in more detail.

\subsection{Extracting Features Describing App Stores}
\label{sec:rq1Methodology}

Our basic assumption is that \AppAstores{} can be categorized based on a finite set of \tbdWdims.
The features would correspond to traits of the app store where they describe the distinguishing qualities or functional characteristics of the app store.
We encode these \tbdWdims as binary values, i.e., each store \emph{has} or \emph{does not have} a given feature.

In order to identify such \tbdWdims, we first created a seeding set of representative app stores.
We started by enumerating well-known app stores that we were aware of (\step{1}).
Once this set of representative app stores was created, we used an iterative process to identify the \tbdWdims that we felt best characterized these stores (\step{2}).
We then used these \tbdWdims to describe each store.

\subsubsection{Stage 1: Identifying \tbdWDims}

First, each of the six authors was tasked with identifying representative characteristics of five stores and the possible \tbdWdims for each.
Each author worked alone in this step; however, to seek better reliability as well as encourage diverse opinions, each store was assigned to two authors.
We list the 15 stores that were assigned in this step with a short description in \Cref{tab:feature-ext-stores}.
After that, all of the authors met as a group to discuss their findings and further refine the proposed \tbdWdim set.

In the subsequent iterations, the authors worked in pairs, and the pairings were reassigned after each iteration (\step{2}).
In these iterations, each author-pair was assigned a set of 2--3 app stores and was asked to describe them using the current set of \tbdWdims; a key concern was to evaluate whether the existing \tbdWdims were sufficient or needed refinement.
For each store, each author-pair analyzed both its store-front and its documentation; in some cases, we could navigate the store as users but not as developers, in these cases, we relied on the store's supporting documentation.

After this step, the six authors discussed their findings as a group and updated the set of \tbdWdims.
The \tbdWdims were discussed in detail to ensure that they were conceptually independent from each other.
We also made sure that each \tbdWdim applied to at least one store to ensure that it was relevant.

Our process leveraged ideas from the coding process of \emph{Grounded theory}~\cite{walker2006grounded} to extract the \tbdWdims of app stores, and followed the practice of open card sorting~\cite{coxon1999sorting} to create the categorized \tbdWdim set.
Similar to prior work\cite{adolph2011using,hoda2012developing,masood2020agile}, we followed practices of \emph{Grounded theory}'s coding process to extract the \tbdWdims --- where we consider codes as a specific feature of app store operation --- and stopped when we reached saturation with no new \tbdWdims added after a new round of describing app stores.
Similar to prior work\cite{vassallo2020developers,chen2021maintenance,wang2022demystifying}, we applied card sorting to the collected \tbdWdims so inter-related \tbdWdims are grouped together.
The authors formed a group in this process and discussed how different \tbdWdims belong to the same conceptual group and stopped when consensus was reached.

\begin{table*}[ht]
  \centering
  \footnotesize
  \caption{Investigated stores for feature extraction}
  \newcommand{\wzIndentWidth}{2mm}
  \label{tab:feature-ext-stores}
  \begin{adjustbox}{max width=\textwidth}
  \begin{tabular}{@{}p{90pt}p{240pt}@{}}
    \toprule
    Store & Description \\
    \midrule
    Google Play Store & Google's app store for Android \\
    Apple App Store & Store for Apple devices \\
    Samsung GalaxyApps & Store specifically for Samsung devices\\
    GitHub Marketplace & Providing applications and services to integrate with GitHub platform \\
    Atlassian Marketplace & Providing applications and services to integrate with various Atlassian products \\
    Homebrew & Package manager for MacOS \\
    MacPorts & A package manager for MacOS\\
    Ubuntu Packages & Software repository for the Ubuntu Linux distribution, with a official front end Ubuntu Software Center \\
    Steam & Gaming focused app store running on multiple operating systems (e.g., Windows, Linux) \\
    Nintendo EShop & Provides applications for Nintendo devices (e.g., Nintendo Switch, Nintendo 3DS)\\
    GoG & Gaming focused store focusing on providing DRM free games \\
    JetBrains Plugin Store & Provides plugins to enhance the behavior of JetBrains IDEs \\
    VSCode Marketplace & Provides plugins to enhance the editor \\
    Chrome Web Store & Provides extensions to enhance Chromium based web browsers \\
    AWS Marketplace & Provides servers and cloud services \\
    \bottomrule
  \end{tabular}
\end{adjustbox}
\end{table*}

\subsubsection{Stage 2: Expanding Our Set of App Stores and Further Evaluation and Refinement the \tbdWDims}
\label{sec:methodology-stage-2}

Once we had agreed on the \tbdWdims, our next goal was to verify that these \tbdWdims were capable of describing other app stores that were not part of the initial seed, or if \tbdWdims were missing or needed refinement.
We used a common search engine, \emph{Google}, to expand our set of app stores in a methodical manner (\step{3}).
To achieve the goal of including a broad range of yet undiscovered app stores, we first derived general search terms by combining synonyms for "app" and "store".
More specifically, we have built all possible combinations of the following terms to construct our search queries:

\begin{description}
  \item[First half] \texttt{software}, \texttt{(extension -hair -lash)}, \texttt{(addon OR add-on)},\\
    \texttt{solution}, \texttt{plugin OR plug-in}, \texttt{install}, \texttt{app}, \texttt{package}
  \item[Second half] \texttt{repository}, \texttt{shop}, \texttt{("app store" OR store)}, \texttt{("market place" OR marketplace)}, \texttt{manager}
\end{description}

For example, a concrete query was created by combining \texttt{app} and \texttt{("app store" OR store)}.
For some queries, it was necessary to refine the term to avoid noise in the results; for example, searching for the term \texttt{extension} would mainly return results related to hair product or eye lashes.
In total, with 8 synonyms for \texttt{app} and 5 synonyms for \texttt{store} we were able to create 40 unique \emph{Google} search queries.
We felt confident that these search terms were representative when we found that the initial seed list had been exhaustively covered.

\smallskip
Our \emph{Google} search was performed in November 2020.
We queried and stored the search results for each search query.
Two authors classified each result as to whether or not it corresponded to an app store.
We devised two inclusion criteria for this decision: 1) the store in question should offer software or software-based services, and 2) the store in question should offer an end-to-end experience for users (ordering, delivery, installation).
We considered only direct hits to the store (e.g., product page), and we explicitly excluded results that contain only indirect references to a store, such as blog posts, videos, or news.
Any disagreements were resolved through discussion.
However, despite our initial effort of maintaining a clear set of inclusion criteria for app stores, several corner cases became apparent during the labeling process.
The first two authors discussed these cases as they arose, and continually updated the inclusion criteria throughout the labeling process.
In a few special cases no agreement could be reached, so another author acted as a moderator and resolved the disagreement by a majority vote.
Over time, the inclusion criteria and \tbdWdims evolved and eventually reached a stable state (in \step{3}). Our final state of the inclusion/exclusion criteria is presented as the working definition for app stores defined in \Cref{sec:app-store-def}.

The classification of search results was stopped when a new results page did not contain any new links to app stores, or once all 10 retrieved pages were analyzed.
Initially, 586 URLs were examined by the first two authors until a saturation of agreement was reached (90.7\% agreement rate).
The first author continued to label the rest.
In the end, a total of 1,600 URLs were labeled.
Multiple search results can refer to the same store; these duplicates were detected and eliminated by using the root domain of the URL.
The most common duplicate references were found for the domains \domain{google.com} (61), \domain{apple.com} (22), and \domain{microsoft.com} (18).
In the end, we found 291 stores.
We note that the exact number of unique stores may differ since two root domains can point to the same store, \domain{kodi.tv} and \domain{kodi.wiki}, or the same root domain may contain multiple stores, \domain{chrome.google.com} and \domain{play.google.com}.

In the next step (\step{5}), we constructed and labelled a set of app stores based on our identified features from \step{2}.
We began from the URLs labelled in the last step and selected the first three occurring stores for each search term; this resulted in 104 URLs pointing to 48 unique stores.
Two of the stores were could not be accessed by the authors: \name{ASRock App Shop} requires physical hardware to use it, and \name{PLCnext Store}'s website was unresponsive at the time of labelling.  
These stores were removed from the list.
In addition, we discussed several more stores that we felt deserved explicit investigation: \name{AWS}, \name{Flatpak}, \name{GoG}, \name{MacPorts}, \name{Nintendo eShop}, \name{Steam}, and Samsung's \name{Galaxy Store}.
These are the stores that the authors investigated in \step{2} but did not show up in the first three occurring results from the search terms.
Meanwhile, the added stores all show up in the list of 291 stores identified by all labelled URLs.

We thus selected and labeled a total of \tbdVlabelledStores{} app stores.
This sample is non-exhaustive, but we believe that our wide range of search queries has created a representative sample of the population of app stores that enables our experiments.

The first two authors proceeded to describe 12 app stores, selected as the first from each search query, using the set of \tbdWdims.
This was done to make sure there was consistency in the interpretation and use of each \tbdWdim.
After that, the first author labelled the remaining stores.

To check the applicability of our dimensions and the labelling guidelines, we have measured the inter-rater agreement between two authors on the 12 stores.
We used the Cohen's Kappa~\cite{cohen1960coefficient} as a measurement for our inter-rater agreement.
The Cohen's Kappa is widely used in software engineering research~\cite{perez2020systematic}.
We have reached an agreement of 86.3\% with Cohen's Kappa~\cite{cohen1960coefficient} of 0.711).
Our agreement based on the Cohen's Kappa is considered as a substantial~\cite{lantz1996behavior} inter-rater agreement suggesting a high confidence of agreement between the two raters.

The outcomes of RQ1 were a list of \tbdWdims that describe the main characteristics of app stores grouped by dimensions, and a set of \tbdVlabelledStores{} \AppStores, each labelled using these \tbdWdims.

\subsection{Finding Natural Groupings of App Stores}
\label{sec:rq2Methodology}

With the outcomes of RQ1, we next performed a \kmeans clustering analysis to identify groups of similar stores.
\kmeans is a popular clustering algorithm widely used in software engineering research~\cite{pickerill2020phantom,khatibi2014flexible,al2019empirical,kuchta2018correctness}.
It groups vectorized data points iteratively until $k$ centroids are formed.
We used the \emph{K-means++} implementation~\cite{arthur2006k} to conduct the clustering process.

\subsubsection{Stage 3: Cluster Analysis}

To identify related app stores, we decided to cluster them using the \kmeans algorithm (\step{6}).

To prepare our labels for the \kmeans clustering process, we converted each
label of the feature to a binary value: 1 if the store has the feature, and
0 if it does not.
Having binary-encoded data ensured that we do not suffer from having categorical values that do not make sense in the scope of \kmeans.
However, performing \kmeans on binary data can also be problematic, since the initial centroids selected will be binary.
To mitigate this issue, we applied Principal Component Analysis (PCA)~\cite{wold1987principal} to both reduce the dimensional space and to produce a mapping in the continuous range.
We kept all principal components that explained a variance of at least 0.05.
Finally, we used the Silhouette method~\cite{rousseeuw1987silhouettes} to determine the best number of clusters within a range of 1 to 20.
To identify the \tbdWdims that best characterize each cluster, we have calculated the deviation of each cluster \emph{centroid} (i.e, the \emph{center} of the cluster) from the \emph{centroid-of-centroids} (\emph{C}) over all clusters.

As an unsupervised method, the result of \kmeans provides only the clustering result with the stores in each cluster.
We then further discussed the results of the \kmeans process and categorized the clusters by the properties of the contained stores.
Following our discussion and categorization, we assigned groupings and names to each of the clusters.

\section{Results}
\label{sec:results}

In this section, we present the results of our investigations into each of the research questions.
The results are organized based on the three stages discussed in \Cref{sec:methodology}.

\subsection*{\tbdSrqone}\label{sec:rq1}

\subsubsection*{Stage 1: \tbdWDims characterizing app stores}

As discussed in \Cref{sec:rq1Methodology}, we derive a set of \tbdWdims and organizational categories that describe the set of studied app stores; the results of these efforts are summarized in \Cref{tab:dim_listing}.
We have modelled the \tbdWdims as a binary representation; thus, each store either has or does not have this \tbdWdim.
We note that for some categories, the \tbdWdims are mutually exclusive; for example, in the category \emph{Rights Management}, a store can have either \emph{Creator managed DRM} or \emph{Store-enforced DRM}, but not both.
In other categories, an app store may have several of the \tbdWdims within a given category; for example, there may be several kinds of communication channels between users, app creators, and the store owner for a given app store.
We now describe each high-level category in detail.

\begin{table*}
	\scriptsize
	\newcommand{\wzIndentWidth}{2mm}
	\setlength\extrarowheight{1pt}
	\centering
	\caption{Features for describing app stores}
	\label{tab:dim_listing}
  \begin{adjustbox}{max width=\textwidth}
  \begin{tabular}{@{}p{60mm}p{110mm}@{}}
		\toprule
		Feature                                                                 & Description                                                                                                                                         \\
		\midrule
		\textit{\textbf{Monetization}}                                          &
		The type of payment options directly offered by the app store.
		\\
		\hspace{\wzIndentWidth} Free                                            & Free as in in the product can be directly acquired                                                                                                  \\
		\hspace{\wzIndentWidth} One-time payment                                & A single payment needed for the product                                                                                                             \\
		\hspace{\wzIndentWidth} Seat-based subscription                         & The subscription is based on the number of products provided                                                                                        \\
		\hspace{\wzIndentWidth} Time-based subscription                         & A payment is needed by a set time interval (e.g.,, monthy, yearly)                                                                                  \\
		\hspace{\wzIndentWidth} Resourced-based subscription                    & A payment is needed by the amount of resource used (e.g., API calls, CPU time)                                                                      \\
		\hspace{\wzIndentWidth} Micro-transaction                               & Additional payment can be collected based on additional feature offered in a product                                                                \\
		\hspace{\wzIndentWidth} Custom pricing (i.e., ``Contact us for price'') &
		The actual price is based on a per case situation; this happens mostly in business-focused app stores                                                                                                                         \\
		\textit{\textbf{Rights Management*}}                                    & How does the store take care of DRM on the product provided.                                                                                        \\
		\hspace{\wzIndentWidth} Creator-managed DRM                             & No DRM is offered by the store and is taken care of by the creator                                                                                  \\
		\hspace{\wzIndentWidth} Store-enforced DRM                              & Store wide DRM for every product offered in the store                                                                                               \\
		\textit{\textbf{Do I need an account?*}}                                & Whether it is possible to use the app store without registration.                                                                                   \\
		\hspace{\wzIndentWidth} Account required                                & An account is required to use the store                                                                                                             \\
		\hspace{\wzIndentWidth} No registration possible                        & The store does not have an account system                                                                                                           \\
		\hspace{\wzIndentWidth} Some features requires registration             & Some content of the store is locked behind an account, but the store can be used without one.                                                       \\
		\textit{\textbf{Product type}}                                          & The type of product the store offers.                                                                                                               \\
		\hspace{\wzIndentWidth} Standalone apps                                 & The product operates by itself                                                                                                                      \\
		\hspace{\wzIndentWidth} Extension/add-ons to apps/hardware              & The product acts as a feature extension to another application/hardware                                                                             \\
		\hspace{\wzIndentWidth} Service/resources                               & The software product is a service                                                                                                                   \\
		\hspace{\wzIndentWidth} Package/library                                 & The product is not an end-user product, but offers functionality to other products                                                                  \\
		\textit{\textbf{Target audience*}}                                      & The intended users of the app store.                                                                                                                \\
		\hspace{\wzIndentWidth} General purpose                                 & The app store is intended to be used by everyone.                                                                                                   \\
		\hspace{\wzIndentWidth} Domain-specific                                 & The app store have a specific focus and is very unlikely to be used by a normal person                                                              \\
		\textit{\textbf{Type of product creators}}                              & The type of creators who submits products to the app store.                                                                                         \\
		\hspace{\wzIndentWidth} Business                                        & The creators mostly have a commercial or business
		focus                                                                                                                                                                                                                         \\
\hspace{\wzIndentWidth} Community                                       & The creators are from the community (e.g., open source developers)                                                                                  \\
		\textit{\textbf{Intent of app store}}                                   & The reason why the app store exists from the app stores' perspective.                                                                               \\
		\hspace{\wzIndentWidth} Community building/support                      & The app store aims to serve a technical community                                                                                                   \\
		\hspace{\wzIndentWidth} Profit                                          & The app store aims to earn money                                                                                                                    \\
		\hspace{\wzIndentWidth} Centralization of product delivery              & The app store aims to provide a way for customer to gather apps in a centralized way                                                                \\
		\hspace{\wzIndentWidth} Expanding a platform popularity/usefulness      & The app store aims to extend functionality from the platform it is based on                                                                         \\
		\textit{\textbf{Role of intermediary}}                                  & The role app store play between the creator of products and the customer of the app store.                                                          \\
		\hspace{\wzIndentWidth} Embedded advertisement API                      & Provides an advertisement method for creators to take advantage of                                                                                  \\
		\hspace{\wzIndentWidth} CI/CD                                           & Offers continuous integration/continuous deployment for creators                                                                                    \\
		\hspace{\wzIndentWidth} Checks at run time                              & Provide checks when apps installed from the app store is ran                                                                                        \\
		\hspace{\wzIndentWidth} Checks before making available to the customer  & Provide checks when an app is submitted to the app store for quality reasons                                                                        \\
		\textit{\textbf{Composability*}}                                        & The relationship between products provided in the app store.                                                                                        \\
		\hspace{\wzIndentWidth} Independent                                     & The products in the app store are unrelated to each other                                                                                           \\
		\hspace{\wzIndentWidth} Vendor internal add-on/extension/unlock         & Some products can be based on other products from the same creator (e.g., game DLC, app feature packs)                                              \\
		\hspace{\wzIndentWidth} Package manager type of app relationship        & A dependency relationship exists between products in the app store                                                                                  \\
		\textit{\textbf{Analytics}}                                             & The type of analytical data provided by the app store.                                                                                              \\
		\hspace{\wzIndentWidth} Sentiment and popularity ratings                & Information related to the popularity of a product (e.g., downloads, score ratings)                                                                 \\
		\hspace{\wzIndentWidth} Marketing feedback                              & Information related to marketing for the creator (e.g., sales, conversion, retention)                                                               \\
		\hspace{\wzIndentWidth} Product usage data                              & Information related to the usage of the product. (e.g., logging, user profiling)                                                                    \\
		\textit{\textbf{Communication channels}}                                & The methods where different parties of the app store can communicate with each other.                                                               \\
		\hspace{\wzIndentWidth} Documentation                                   & Information related to the operation of the store (e.g., instructions to install applications)                                                      \\
		\hspace{\wzIndentWidth} Product homepage                                & A homepage for a specific product in the app store                                                                                                  \\
		\hspace{\wzIndentWidth} Ratings                                         & Any form of rating customers can give to a product (e.g., star, score, up/down vote)                                                                \\
		\hspace{\wzIndentWidth} Written reviews (in text)                       & A written viewer where customers can write their experience to the product.                                                                         \\
		\hspace{\wzIndentWidth} Community forum                                 & A forum like feature offered by the store where people can discuss things related to the store/product.                                             \\
		\hspace{\wzIndentWidth} Support ticket                                  & A system where customers can inquiry for support questions related to the product offered by the store.                                             \\
		\hspace{\wzIndentWidth} Promotion/marketing                             & The store offers a way to provide promotional/marketing feature to the products in the app store (e.g., featured apps, top downloads of the month). \\
		\bottomrule
	\end{tabular}
  \end{adjustbox}

	\bigskip

	\emph{*:} Categorical values are mutually exclusive; one and only one
	categorical value in the dimension can apply to a given store.

\end{table*}

\paragraph{Monetization} describes what, if any, payment options are provided to the user directly by the store.
If a product is offered free within the store, but requires an activation key obtained elsewhere, we consider that the product is free.
While most of the options are self-explanatory, some may be less obvious.
For example, \name{GitHub Marketplace} offers \emph{seat-based subscriptions} where app pricing is calculated by the number of installations made to individual machines; usually, this occurs within the context of enterprise purchase.
Also, \name{AWS} offers \emph{resource-based subscription} where the price charged is determined by the amount of resources --- such as cloud storage and CPU time --- that are used during the execution of the service.

\paragraph{Rights Management} describes the Digital Rights Management (DRM) policy of the store; the values describe whether the store uses a store-wide DRM feature.
For example, for \name{Steam}, all games have DRM encryption, whereas the \name{F-Droid} store contains only open source apps, so there is no need for DRM.

\paragraph{Do I need an account?} describes whether a user can access and use the store without being registered with the app store.
We find that most stores are either \emph{account required} (e.g., \appleAppStore) or \emph{no registration possible} (e.g., \name{Snapcraft}).
However, we also found that some stores can be used without an account for some purposes, with other features requiring explicit registration; for example, the \name{Microsoft Store} allows users to download free applications without an account, but to purchase an app or leave a review, an account is required.

\paragraph{Product type} describes the kinds of applications that are offered by the store.
For example, \googlePlay and \name{Steam} focus  on \emph{standalone apps}, the \name{VSCode Marketplace} store offers \emph{add-ons} to an existing tool, and \name{AWS} allows users to ``rent'' web-based \emph{resources and services}.

\paragraph{Target audience} describes the intended user base of the store.
\emph{General-purpose} stores offer products aimed at the broad general public of everyday technology users; this includes stores such as \googlePlay, \name{Steam}, and the \name{Chrome Web Store}.
\emph{Domain-specific} stores, on the other hand, have a dedicated focus on a specialized field; for example, \name{Adobe Magento} focuses on building e-commerce platforms.

\paragraph{Type of product creators} describes the typical focus of creators submitting applications to the store.
We distinguish between two groups of creators: those with a commercial or business focus, and those with community focus such as open source developers.

\paragraph{Intent of app store} describes the perceived high-level goals of the app store.
The values are derived from the app stores' own descriptions of their goals,  often found in \emph{``About us''} web pages.
For example, both \name{F-Droid} and \name{ApkPure} are \android app
stores; however, \name{F-Droid}'s focus is to provide a location to
download and support FOSS software, while \name{ApkPure}'s goal is to
provide a location for users to be able to download \android apps when \googlePlay may be unavailable.

\paragraph{Role of intermediary} describes the roles that the app store plays in mediating between the users and creators; these are software engineering-related services that are mostly independent of each other.
For example, \emph{checks at run time} tracks if the app store ensures that its products function correctly (e.g., \name{Steam} tracking game stats).
Also, \emph{CI/CD} indicates that the app store provides explicit support for continuous integration and deployment of the apps, which may be linked to specific development tools used by the creator.

\paragraph{Composability} describes the relationship between products offered by the store.
App stores of \emph{independent} composability offer products that have no relationship with each other, such as  \name{Firefox Add-ons}.
\emph{Vendor internal add-on/extension/unlock} means that the products within the app store can be based on each other, but only when they are from the same vendor, such as game DLC and micro-transaction unlocks.
\emph{Package managers} contain apps that can have complicated dependency relationships regardless of the creator of the products, such as the \emph{Ubuntu} package management tool \name{apt}.

\paragraph{Analytics} describes what kind of diagnostic information is provided by the store.
We distinguish between three kinds:
\emph{Sentiment and popularity ratings} offer user-based information related to store products, such as number of installs in \name{Home Assistant}.
\emph{Marketing feedback} tracks telemetry information for creators on the performance of their product, such as \name{GitHub Marketplace} tracking retention rate for their products for creators.
\emph{Product usage data} details the observed usage of the products; for example, \name{Steam} tracks the average number of hours users spend on each product.

\paragraph{Communication channels} tracks the types of methods the store directly offers for communications between both users and creators.
Since most stores offer a \emph{product homepage} for each of their products, the app creators are largely free to put any information here.
This means that if a creator wishes, they can put links to other communication methods external to the store.
We do not track such information here since it is product dependent instead of store dependent.
While ratings and reviews/comments are often paired together, during our exploration, we found cases where user ratings were permitted but user reviews were not; thus, we have separate values for ratings and reviews.
Communication channels can take various forms with different variability, for example, some stores allow responses for reviews.
For this aspect, we stay at a high level based on the functionality of the communication channels and consider the variations as detailed implementation for each functionality.

\subsubsection*{Stage 2: Expanded collection of app stores and labelled set of representative stores}\label{sec:data-collection}

In stage 1, we identified \tbdVlabelledStores{} store candidates.
To provide the required data for our experiments, two authors explored these stores to identify which of the \emph{fundamental \tbdWdims} of the previous stage are true for each store.
The query results are summarized in \Cref{tab:query_table}, where we list the search term construction keywords and the first 3 occurrence of stores by the search term.
For example, in search term constructed from \texttt{(addon OR add-on)} and \texttt{("market place") OR marketplace}, the first 3 occurrences are \googlePlay, \name{PrestaShop}, and \name{CS-Cart}.

\begin{hassanbox}{\columnwidth}[4ex]
	There are many app stores beyond \googlePlay and \appleAppStore.  These app stores exhibit a diverse set of features.
\end{hassanbox}

\newcommand{\na}{{\color{black!33}$\phi$}}

\begin{sidewaystable*}[htbp]
\scriptsize
\centering
\caption{First three identified stores for each \emph{Google} query}
\label{tab:query_table}
\begin{tabular}{@{}p{32mm}p{30mm}p{30mm}p{25mm}p{25mm}p{25mm}@{}}
\toprule
                               & {\tt ("app store" OR store)}                   & {\tt ("market place" OR marketplace)}                 & {\tt shop}                                    & {\tt repository}                       & {manager}                   \\
\midrule
{\tt app}                      & Apple App Store, Google Play                   & BigCommerce, Google Play, HubSpot                     & Apple App Store                               & F-Droid, Guardian Project, IzzyOnDroid & Google Play                 \\
{\tt software}                 & Mac App Store                                  & MarketPlaceKit, Sellacious, CS-Cart                   & \na                                           & \na                                    & \na                         \\
{\tt (addon OR add-on)}        & Mac App Store, Home Assistant, Firefox Add-ons & Google Play, PrestaShop, CS-Cart                      & PrestaShop, Chrome Web Store                  & Kodi                                   & CurseForge, Ajour, Minion   \\
{\tt (plugin OR plug-in)}      & Google Play, SketchUcation, RICOH THETA        & WordPress, JetBrains                                  & Bukkit, Plugin Boutique                       & WordPress, JetBrains                   & Jenkins, JMeter, Autodesk   \\
{\tt (extension -hair -lash)}  & Chrome Web Store, Microsoft Edge               & VSCode Marketplace, Adobe Magento, Chrome Web Store   & Chrome Web Store                              & TYPO3, GNOME SHELL                     & Chrome Web Store            \\
{\tt install}                  & Google Play, Apple App Store                   & Google Play, Eclipse                                  & Apple App Store, Google Play, Microsoft Store & Kodi, Home Assistant, DockerHub        & Google Play, APKPure, Daz3D \\
{\tt solution}                 & Mac App Store, Microsoft Store                 & CS-Cart                                               & \na                                           & \na                                    & \na                         \\
{\tt (software library -book)} & Microsoft Store                                & VSCode Marketplace, QT Extensions, GitHub Marketplace & \na                                           & \na                                    & \na                         \\
{\tt package}                  & Apple App Store, Google Play, Snapcraft        & CS-Cart, concrete5                                    & Google Play                                   & Packagist, PyPI, Ubuntu Packages       & Chocolatey, NPM, NuGet      \\
\bottomrule
\end{tabular}
\end{sidewaystable*}

\subsection*{\tbdSrqtwo} \label{sec:data-analysis}

Using the labeled data of the \tbdVlabelledStores{} stores, we were able to perform the \kmeans cluster analysis that we have introduced in \Cref{sec:rq2Methodology}.
With the number of clusters guided by the Silhouette method to choose the best $k$ value for \kmeans, our clustering resulted in eight clusters.

Due to the nature of unsupervised methods, \kmeans is able to identify only the clusters and their members; no real-world meanings are extracted for why the cluster members belong together.
It is also important to note that the \kmeans algorithm performs \emph{hard clustering}; that is, it creates a partitioning of the stores into mutually exclusive groups that together span the whole space.
Thus each store will be assigned to the unique cluster that the algorithm considers to best represent it.
For this reason, the raw results from \kmeans should not be seen as authoritative, but rather as a vehicle for identifying groups of stores with similar characteristics.
Therefore, we leverage the \kmeans clustering and further examine the clusters in detail to try to derive a human understandable categorization of the stores.

We start by analyzing the differences between clusters by analyzing the definitive characteristics of each cluster.
In \Cref{table:centroid_deviations}, we show the details of the top 10 \tbdWdims that deviate the most from the \emph{C}.
Column C contains the \emph{centroid-of-centroids} with values for each \tbdWdim.
The remaining columns represent each cluster by an index from 1 to 8.
The values in these columns represent the proportion of app stores in the cluster with a specific \tbdWdim, the mean, and the background color of each cell represent the deviation of the particular cluster \emph{centroid} (i.e., difference between the centroid of this cluster and the centroid-of-centroids for the \tbdWdim).
Each row corresponds to a \tbdWdim of the stores, which makes it easy to understand which \tbdWdims are descriptive of a cluster.

\newcommand{\attHeader}[1]{\emph{#1}}
\newcommand{\attIndent}{\quad}

\begin{table*}
  \centering
  \scriptsize
  \label{table:centroid_deviations}
  \caption{The \tbdVbestK{} clusters found by the \emph{K-means} algorithm,
    with top deviated \tbdWdims from the centroid of centroids (C).
Each cell with a value represents one of the ten most influential features of the corresponding cluster. The number indicates the percentage of stores with the specific feature. The color encodes whether stores in that cluster are less (magenta) or more (green) likely to have the feature, compared to the centroid.}
    \vspace{5pt}

    \begin{adjustbox}{max width=\textwidth}
    \begin{tabular}{@{}lrcccccccc}
    \toprule
 &  & \multicolumn{8}{@{}c@{}}{\bf Cluster Index} \\
 \cmidrule(l){3-10}
{\bf Features} & \bf{C} & 1 & 2 & 3 & 4 & 5 & 6 & 7 & 8 \\
 \midrule
\attHeader{Monetization} & & & & & & & & & \\
{\attIndent} Free & 1.00 & {\cellcolor[HTML]{000000}} \color[HTML]{F1F1F1} \color{white} {\cellcolor{white}}  & {\cellcolor[HTML]{000000}} \color[HTML]{F1F1F1} \color{white} {\cellcolor{white}}  & {\cellcolor[HTML]{000000}} \color[HTML]{F1F1F1} \color{white} {\cellcolor{white}}  & {\cellcolor[HTML]{000000}} \color[HTML]{F1F1F1} \color{white} {\cellcolor{white}}  & {\cellcolor[HTML]{000000}} \color[HTML]{F1F1F1} \color{white} {\cellcolor{white}}  & {\cellcolor[HTML]{000000}} \color[HTML]{F1F1F1} \color{white} {\cellcolor{white}}  & {\cellcolor[HTML]{000000}} \color[HTML]{F1F1F1} \color{white} {\cellcolor{white}}  & {\cellcolor[HTML]{000000}} \color[HTML]{F1F1F1} \color{white} {\cellcolor{white}}  \\
{\attIndent} One-time payment & 0.35 & {\cellcolor[HTML]{F4C1DF}} \color[HTML]{000000} 0.00 & {\cellcolor[HTML]{000000}} \color[HTML]{F1F1F1} \color{white} {\cellcolor{white}}  & {\cellcolor[HTML]{F4C1DF}} \color[HTML]{000000} 0.00 & {\cellcolor[HTML]{F4C1DF}} \color[HTML]{000000} 0.00 & {\cellcolor[HTML]{000000}} \color[HTML]{F1F1F1} \color{white} {\cellcolor{white}}  & {\cellcolor[HTML]{000000}} \color[HTML]{F1F1F1} \color{white} {\cellcolor{white}}  & {\cellcolor[HTML]{000000}} \color[HTML]{F1F1F1} \color{white} {\cellcolor{white}}  & {\cellcolor[HTML]{71B038}} \color[HTML]{F1F1F1} 1.00 \\
{\attIndent} Seat-based subscription & 0.09 & {\cellcolor[HTML]{000000}} \color[HTML]{F1F1F1} \color{white} {\cellcolor{white}}  & {\cellcolor[HTML]{B5DF82}} \color[HTML]{000000} 0.50 & {\cellcolor[HTML]{000000}} \color[HTML]{F1F1F1} \color{white} {\cellcolor{white}}  & {\cellcolor[HTML]{000000}} \color[HTML]{F1F1F1} \color{white} {\cellcolor{white}}  & {\cellcolor[HTML]{000000}} \color[HTML]{F1F1F1} \color{white} {\cellcolor{white}}  & {\cellcolor[HTML]{000000}} \color[HTML]{F1F1F1} \color{white} {\cellcolor{white}}  & {\cellcolor[HTML]{000000}} \color[HTML]{F1F1F1} \color{white} {\cellcolor{white}}  & {\cellcolor[HTML]{000000}} \color[HTML]{F1F1F1} \color{white} {\cellcolor{white}}  \\
{\attIndent} Time-based subscription & 0.30 & {\cellcolor[HTML]{000000}} \color[HTML]{F1F1F1} \color{white} {\cellcolor{white}}  & {\cellcolor[HTML]{A9D874}} \color[HTML]{000000} 0.75 & {\cellcolor[HTML]{F7CBE4}} \color[HTML]{000000} 0.00 & {\cellcolor[HTML]{000000}} \color[HTML]{F1F1F1} \color{white} {\cellcolor{white}}  & {\cellcolor[HTML]{000000}} \color[HTML]{F1F1F1} \color{white} {\cellcolor{white}}  & {\cellcolor[HTML]{000000}} \color[HTML]{F1F1F1} \color{white} {\cellcolor{white}}  & {\cellcolor[HTML]{000000}} \color[HTML]{F1F1F1} \color{white} {\cellcolor{white}}  & {\cellcolor[HTML]{8AC34F}} \color[HTML]{000000} 0.86 \\
{\attIndent} Resource-based subscription & 0.05 & {\cellcolor[HTML]{000000}} \color[HTML]{F1F1F1} \color{white} {\cellcolor{white}}  & {\cellcolor[HTML]{000000}} \color[HTML]{F1F1F1} \color{white} {\cellcolor{white}}  & {\cellcolor[HTML]{000000}} \color[HTML]{F1F1F1} \color{white} {\cellcolor{white}}  & {\cellcolor[HTML]{000000}} \color[HTML]{F1F1F1} \color{white} {\cellcolor{white}}  & {\cellcolor[HTML]{000000}} \color[HTML]{F1F1F1} \color{white} {\cellcolor{white}}  & {\cellcolor[HTML]{000000}} \color[HTML]{F1F1F1} \color{white} {\cellcolor{white}}  & {\cellcolor[HTML]{000000}} \color[HTML]{F1F1F1} \color{white} {\cellcolor{white}}  & {\cellcolor[HTML]{000000}} \color[HTML]{F1F1F1} \color{white} {\cellcolor{white}}  \\
{\attIndent} Micro-transactions & 0.11 & {\cellcolor[HTML]{000000}} \color[HTML]{F1F1F1} \color{white} {\cellcolor{white}}  & {\cellcolor[HTML]{000000}} \color[HTML]{F1F1F1} \color{white} {\cellcolor{white}}  & {\cellcolor[HTML]{000000}} \color[HTML]{F1F1F1} \color{white} {\cellcolor{white}}  & {\cellcolor[HTML]{000000}} \color[HTML]{F1F1F1} \color{white} {\cellcolor{white}}  & {\cellcolor[HTML]{000000}} \color[HTML]{F1F1F1} \color{white} {\cellcolor{white}}  & {\cellcolor[HTML]{000000}} \color[HTML]{F1F1F1} \color{white} {\cellcolor{white}}  & {\cellcolor[HTML]{000000}} \color[HTML]{F1F1F1} \color{white} {\cellcolor{white}}  & {\cellcolor[HTML]{589B28}} \color[HTML]{F1F1F1} 0.86 \\
{\attIndent} Custom Pricing & 0.01 & {\cellcolor[HTML]{000000}} \color[HTML]{F1F1F1} \color{white} {\cellcolor{white}}  & {\cellcolor[HTML]{000000}} \color[HTML]{F1F1F1} \color{white} {\cellcolor{white}}  & {\cellcolor[HTML]{000000}} \color[HTML]{F1F1F1} \color{white} {\cellcolor{white}}  & {\cellcolor[HTML]{000000}} \color[HTML]{F1F1F1} \color{white} {\cellcolor{white}}  & {\cellcolor[HTML]{000000}} \color[HTML]{F1F1F1} \color{white} {\cellcolor{white}}  & {\cellcolor[HTML]{000000}} \color[HTML]{F1F1F1} \color{white} {\cellcolor{white}}  & {\cellcolor[HTML]{000000}} \color[HTML]{F1F1F1} \color{white} {\cellcolor{white}}  & {\cellcolor[HTML]{000000}} \color[HTML]{F1F1F1} \color{white} {\cellcolor{white}}  \\
\attHeader{Rights Management} & & & & & & & & &\\
{\attIndent} Creator managed DRM & 0.72 & {\cellcolor[HTML]{000000}} \color[HTML]{F1F1F1} \color{white} {\cellcolor{white}}  & {\cellcolor[HTML]{EA9FCA}} \color[HTML]{000000} 0.25 & {\cellcolor[HTML]{D4EDB3}} \color[HTML]{000000} 1.00 & {\cellcolor[HTML]{000000}} \color[HTML]{F1F1F1} \color{white} {\cellcolor{white}}  & {\cellcolor[HTML]{000000}} \color[HTML]{F1F1F1} \color{white} {\cellcolor{white}}  & {\cellcolor[HTML]{000000}} \color[HTML]{F1F1F1} \color{white} {\cellcolor{white}}  & {\cellcolor[HTML]{000000}} \color[HTML]{F1F1F1} \color{white} {\cellcolor{white}}  & {\cellcolor[HTML]{E07EB3}} \color[HTML]{F1F1F1} 0.14 \\
{\attIndent} Store-enforced DRM & 0.27 & {\cellcolor[HTML]{000000}} \color[HTML]{F1F1F1} \color{white} {\cellcolor{white}}  & {\cellcolor[HTML]{A1D26A}} \color[HTML]{000000} 0.75 & {\cellcolor[HTML]{000000}} \color[HTML]{F1F1F1} \color{white} {\cellcolor{white}}  & {\cellcolor[HTML]{000000}} \color[HTML]{F1F1F1} \color{white} {\cellcolor{white}}  & {\cellcolor[HTML]{000000}} \color[HTML]{F1F1F1} \color{white} {\cellcolor{white}}  & {\cellcolor[HTML]{000000}} \color[HTML]{F1F1F1} \color{white} {\cellcolor{white}}  & {\cellcolor[HTML]{000000}} \color[HTML]{F1F1F1} \color{white} {\cellcolor{white}}  & {\cellcolor[HTML]{81BD44}} \color[HTML]{000000} 0.86 \\
\attHeader{Do I need an account to use the store} & & & & & & & & & \\
{\attIndent} Account Required & 0.33 & {\cellcolor[HTML]{000000}} \color[HTML]{F1F1F1} \color{white} {\cellcolor{white}}  & {\cellcolor[HTML]{000000}} \color[HTML]{F1F1F1} \color{white} {\cellcolor{white}}  & {\cellcolor[HTML]{F5C6E2}} \color[HTML]{000000} 0.00 & {\cellcolor[HTML]{000000}} \color[HTML]{F1F1F1} \color{white} {\cellcolor{white}}  & {\cellcolor[HTML]{B0DC7D}} \color[HTML]{000000} 0.75 & {\cellcolor[HTML]{6BAC34}} \color[HTML]{F1F1F1} 1.00 & {\cellcolor[HTML]{000000}} \color[HTML]{F1F1F1} \color{white} {\cellcolor{white}}  & {\cellcolor[HTML]{91C857}} \color[HTML]{000000} 0.86 \\
{\attIndent} No registration possible & 0.35 & {\cellcolor[HTML]{71B038}} \color[HTML]{F1F1F1} 1.00 & {\cellcolor[HTML]{F4C1DF}} \color[HTML]{000000} 0.00 & {\cellcolor[HTML]{000000}} \color[HTML]{F1F1F1} \color{white} {\cellcolor{white}}  & {\cellcolor[HTML]{F4C1DF}} \color[HTML]{000000} 0.00 & {\cellcolor[HTML]{000000}} \color[HTML]{F1F1F1} \color{white} {\cellcolor{white}}  & {\cellcolor[HTML]{F4C1DF}} \color[HTML]{000000} 0.00 & {\cellcolor[HTML]{71B038}} \color[HTML]{F1F1F1} 1.00 & {\cellcolor[HTML]{000000}} \color[HTML]{F1F1F1} \color{white} {\cellcolor{white}}  \\
{\attIndent} Some features require registration & 0.30 & {\cellcolor[HTML]{000000}} \color[HTML]{F1F1F1} \color{white} {\cellcolor{white}}  & {\cellcolor[HTML]{64A52F}} \color[HTML]{F1F1F1} 1.00 & {\cellcolor[HTML]{000000}} \color[HTML]{F1F1F1} \color{white} {\cellcolor{white}}  & {\cellcolor[HTML]{64A52F}} \color[HTML]{F1F1F1} 1.00 & {\cellcolor[HTML]{000000}} \color[HTML]{F1F1F1} \color{white} {\cellcolor{white}}  & {\cellcolor[HTML]{000000}} \color[HTML]{F1F1F1} \color{white} {\cellcolor{white}}  & {\cellcolor[HTML]{000000}} \color[HTML]{F1F1F1} \color{white} {\cellcolor{white}}  & {\cellcolor[HTML]{000000}} \color[HTML]{F1F1F1} \color{white} {\cellcolor{white}}  \\
\attHeader{Product Type} & & & & & & & & & \\
{\attIndent} Standalone apps & 0.42 & {\cellcolor[HTML]{000000}} \color[HTML]{F1F1F1} \color{white} {\cellcolor{white}}  & {\cellcolor[HTML]{EFB0D6}} \color[HTML]{000000} 0.00 & {\cellcolor[HTML]{000000}} \color[HTML]{F1F1F1} \color{white} {\cellcolor{white}}  & {\cellcolor[HTML]{000000}} \color[HTML]{F1F1F1} \color{white} {\cellcolor{white}}  & {\cellcolor[HTML]{000000}} \color[HTML]{F1F1F1} \color{white} {\cellcolor{white}}  & {\cellcolor[HTML]{000000}} \color[HTML]{F1F1F1} \color{white} {\cellcolor{white}}  & {\cellcolor[HTML]{83BF46}} \color[HTML]{000000} 1.00 & {\cellcolor[HTML]{000000}} \color[HTML]{F1F1F1} \color{white} {\cellcolor{white}}  \\
{\attIndent} Extension/add-ons to apps/hardware & 0.68 & {\cellcolor[HTML]{F4C1DF}} \color[HTML]{000000} 0.33 & {\cellcolor[HTML]{000000}} \color[HTML]{F1F1F1} \color{white} {\cellcolor{white}}  & {\cellcolor[HTML]{000000}} \color[HTML]{F1F1F1} \color{white} {\cellcolor{white}}  & {\cellcolor[HTML]{000000}} \color[HTML]{F1F1F1} \color{white} {\cellcolor{white}}  & {\cellcolor[HTML]{000000}} \color[HTML]{F1F1F1} \color{white} {\cellcolor{white}}  & {\cellcolor[HTML]{000000}} \color[HTML]{F1F1F1} \color{white} {\cellcolor{white}}  & {\cellcolor[HTML]{D34F99}} \color[HTML]{F1F1F1} 0.00 & {\cellcolor[HTML]{000000}} \color[HTML]{F1F1F1} \color{white} {\cellcolor{white}}  \\
{\attIndent} Service/Resources & 0.08 & {\cellcolor[HTML]{000000}} \color[HTML]{F1F1F1} \color{white} {\cellcolor{white}}  & {\cellcolor[HTML]{000000}} \color[HTML]{F1F1F1} \color{white} {\cellcolor{white}}  & {\cellcolor[HTML]{000000}} \color[HTML]{F1F1F1} \color{white} {\cellcolor{white}}  & {\cellcolor[HTML]{000000}} \color[HTML]{F1F1F1} \color{white} {\cellcolor{white}}  & {\cellcolor[HTML]{000000}} \color[HTML]{F1F1F1} \color{white} {\cellcolor{white}}  & {\cellcolor[HTML]{000000}} \color[HTML]{F1F1F1} \color{white} {\cellcolor{white}}  & {\cellcolor[HTML]{000000}} \color[HTML]{F1F1F1} \color{white} {\cellcolor{white}}  & {\cellcolor[HTML]{000000}} \color[HTML]{F1F1F1} \color{white} {\cellcolor{white}}  \\
{\attIndent} Package/Library & 0.17 & {\cellcolor[HTML]{62A32E}} \color[HTML]{F1F1F1} 0.89 & {\cellcolor[HTML]{000000}} \color[HTML]{F1F1F1} \color{white} {\cellcolor{white}}  & {\cellcolor[HTML]{000000}} \color[HTML]{F1F1F1} \color{white} {\cellcolor{white}}  & {\cellcolor[HTML]{000000}} \color[HTML]{F1F1F1} \color{white} {\cellcolor{white}}  & {\cellcolor[HTML]{000000}} \color[HTML]{F1F1F1} \color{white} {\cellcolor{white}}  & {\cellcolor[HTML]{000000}} \color[HTML]{F1F1F1} \color{white} {\cellcolor{white}}  & {\cellcolor[HTML]{000000}} \color[HTML]{F1F1F1} \color{white} {\cellcolor{white}}  & {\cellcolor[HTML]{000000}} \color[HTML]{F1F1F1} \color{white} {\cellcolor{white}}  \\
\attHeader{Target audience} & & & & & & & & & \\
{\attIndent} General purpose & 0.33 & {\cellcolor[HTML]{000000}} \color[HTML]{F1F1F1} \color{white} {\cellcolor{white}}  & {\cellcolor[HTML]{000000}} \color[HTML]{F1F1F1} \color{white} {\cellcolor{white}}  & {\cellcolor[HTML]{F5C4E1}} \color[HTML]{000000} 0.00 & {\cellcolor[HTML]{9ACD61}} \color[HTML]{000000} 0.83 & {\cellcolor[HTML]{000000}} \color[HTML]{F1F1F1} \color{white} {\cellcolor{white}}  & {\cellcolor[HTML]{F5C4E1}} \color[HTML]{000000} 0.00 & {\cellcolor[HTML]{6DAD36}} \color[HTML]{F1F1F1} 1.00 & {\cellcolor[HTML]{000000}} \color[HTML]{F1F1F1} \color{white} {\cellcolor{white}}  \\
{\attIndent} Domain-specific & 0.67 & {\cellcolor[HTML]{000000}} \color[HTML]{F1F1F1} \color{white} {\cellcolor{white}}  & {\cellcolor[HTML]{000000}} \color[HTML]{F1F1F1} \color{white} {\cellcolor{white}}  & {\cellcolor[HTML]{C7E89F}} \color[HTML]{000000} 1.00 & {\cellcolor[HTML]{E795C3}} \color[HTML]{000000} 0.17 & {\cellcolor[HTML]{000000}} \color[HTML]{F1F1F1} \color{white} {\cellcolor{white}}  & {\cellcolor[HTML]{C7E89F}} \color[HTML]{000000} 1.00 & {\cellcolor[HTML]{D5579D}} \color[HTML]{F1F1F1} 0.00 & {\cellcolor[HTML]{000000}} \color[HTML]{F1F1F1} \color{white} {\cellcolor{white}}  \\
\attHeader{Type of product creators} & & & & & & & & & \\
{\attIndent} Business & 0.67 & {\cellcolor[HTML]{ECA6CF}} \color[HTML]{000000} 0.22 & {\cellcolor[HTML]{000000}} \color[HTML]{F1F1F1} \color{white} {\cellcolor{white}}  & {\cellcolor[HTML]{D4539B}} \color[HTML]{F1F1F1} 0.00 & {\cellcolor[HTML]{000000}} \color[HTML]{F1F1F1} \color{white} {\cellcolor{white}}  & {\cellcolor[HTML]{000000}} \color[HTML]{F1F1F1} \color{white} {\cellcolor{white}}  & {\cellcolor[HTML]{C9E8A2}} \color[HTML]{000000} 1.00 & {\cellcolor[HTML]{000000}} \color[HTML]{F1F1F1} \color{white} {\cellcolor{white}}  & {\cellcolor[HTML]{000000}} \color[HTML]{F1F1F1} \color{white} {\cellcolor{white}}  \\
{\attIndent} Community & 0.67 & {\cellcolor[HTML]{000000}} \color[HTML]{F1F1F1} \color{white} {\cellcolor{white}}  & {\cellcolor[HTML]{000000}} \color[HTML]{F1F1F1} \color{white} {\cellcolor{white}}  & {\cellcolor[HTML]{C7E89F}} \color[HTML]{000000} 1.00 & {\cellcolor[HTML]{000000}} \color[HTML]{F1F1F1} \color{white} {\cellcolor{white}}  & {\cellcolor[HTML]{000000}} \color[HTML]{F1F1F1} \color{white} {\cellcolor{white}}  & {\cellcolor[HTML]{E283B7}} \color[HTML]{F1F1F1} 0.11 & {\cellcolor[HTML]{000000}} \color[HTML]{F1F1F1} \color{white} {\cellcolor{white}}  & {\cellcolor[HTML]{000000}} \color[HTML]{F1F1F1} \color{white} {\cellcolor{white}}  \\
\attHeader{Intent of app store} & & & & & & & & & \\
{\attIndent} Community building / support & 0.52 & {\cellcolor[HTML]{000000}} \color[HTML]{F1F1F1} \color{white} {\cellcolor{white}}  & {\cellcolor[HTML]{9ED067}} \color[HTML]{000000} 1.00 & {\cellcolor[HTML]{000000}} \color[HTML]{F1F1F1} \color{white} {\cellcolor{white}}  & {\cellcolor[HTML]{9ED067}} \color[HTML]{000000} 1.00 & {\cellcolor[HTML]{E692C1}} \color[HTML]{000000} 0.00 & {\cellcolor[HTML]{F1B5D9}} \color[HTML]{000000} 0.11 & {\cellcolor[HTML]{000000}} \color[HTML]{F1F1F1} \color{white} {\cellcolor{white}}  & {\cellcolor[HTML]{000000}} \color[HTML]{F1F1F1} \color{white} {\cellcolor{white}}  \\
{\attIndent} Profit & 0.38 & {\cellcolor[HTML]{F2BADC}} \color[HTML]{000000} 0.00 & {\cellcolor[HTML]{000000}} \color[HTML]{F1F1F1} \color{white} {\cellcolor{white}}  & {\cellcolor[HTML]{F2BADC}} \color[HTML]{000000} 0.00 & {\cellcolor[HTML]{F2BADC}} \color[HTML]{000000} 0.00 & {\cellcolor[HTML]{000000}} \color[HTML]{F1F1F1} \color{white} {\cellcolor{white}}  & {\cellcolor[HTML]{B7E085}} \color[HTML]{000000} 0.78 & {\cellcolor[HTML]{F2BADC}} \color[HTML]{000000} 0.00 & {\cellcolor[HTML]{79B73D}} \color[HTML]{F1F1F1} 1.00 \\
{\attIndent} Centralization of product delivery & 0.84 & {\cellcolor[HTML]{000000}} \color[HTML]{F1F1F1} \color{white} {\cellcolor{white}}  & {\cellcolor[HTML]{000000}} \color[HTML]{F1F1F1} \color{white} {\cellcolor{white}}  & {\cellcolor[HTML]{000000}} \color[HTML]{F1F1F1} \color{white} {\cellcolor{white}}  & {\cellcolor[HTML]{000000}} \color[HTML]{F1F1F1} \color{white} {\cellcolor{white}}  & {\cellcolor[HTML]{000000}} \color[HTML]{F1F1F1} \color{white} {\cellcolor{white}}  & {\cellcolor[HTML]{000000}} \color[HTML]{F1F1F1} \color{white} {\cellcolor{white}}  & {\cellcolor[HTML]{000000}} \color[HTML]{F1F1F1} \color{white} {\cellcolor{white}}  & {\cellcolor[HTML]{000000}} \color[HTML]{F1F1F1} \color{white} {\cellcolor{white}}  \\
{\attIndent} Expanding the platform & 0.76 & {\cellcolor[HTML]{000000}} \color[HTML]{F1F1F1} \color{white} {\cellcolor{white}}  & {\cellcolor[HTML]{000000}} \color[HTML]{F1F1F1} \color{white} {\cellcolor{white}}  & {\cellcolor[HTML]{000000}} \color[HTML]{F1F1F1} \color{white} {\cellcolor{white}}  & {\cellcolor[HTML]{000000}} \color[HTML]{F1F1F1} \color{white} {\cellcolor{white}}  & {\cellcolor[HTML]{000000}} \color[HTML]{F1F1F1} \color{white} {\cellcolor{white}}  & {\cellcolor[HTML]{000000}} \color[HTML]{F1F1F1} \color{white} {\cellcolor{white}}  & {\cellcolor[HTML]{DE77AE}} \color[HTML]{F1F1F1} 0.17 & {\cellcolor[HTML]{000000}} \color[HTML]{F1F1F1} \color{white} {\cellcolor{white}}  \\
\attHeader{Role of intermediary} & & & & & & & & & \\
{\attIndent} Embedded Advertisement API & 0.16 & {\cellcolor[HTML]{000000}} \color[HTML]{F1F1F1} \color{white} {\cellcolor{white}}  & {\cellcolor[HTML]{000000}} \color[HTML]{F1F1F1} \color{white} {\cellcolor{white}}  & {\cellcolor[HTML]{000000}} \color[HTML]{F1F1F1} \color{white} {\cellcolor{white}}  & {\cellcolor[HTML]{000000}} \color[HTML]{F1F1F1} \color{white} {\cellcolor{white}}  & {\cellcolor[HTML]{000000}} \color[HTML]{F1F1F1} \color{white} {\cellcolor{white}}  & {\cellcolor[HTML]{000000}} \color[HTML]{F1F1F1} \color{white} {\cellcolor{white}}  & {\cellcolor[HTML]{000000}} \color[HTML]{F1F1F1} \color{white} {\cellcolor{white}}  & {\cellcolor[HTML]{8CC551}} \color[HTML]{000000} 0.71 \\
{\attIndent} CI/CD & 0.05 & {\cellcolor[HTML]{000000}} \color[HTML]{F1F1F1} \color{white} {\cellcolor{white}}  & {\cellcolor[HTML]{000000}} \color[HTML]{F1F1F1} \color{white} {\cellcolor{white}}  & {\cellcolor[HTML]{000000}} \color[HTML]{F1F1F1} \color{white} {\cellcolor{white}}  & {\cellcolor[HTML]{000000}} \color[HTML]{F1F1F1} \color{white} {\cellcolor{white}}  & {\cellcolor[HTML]{000000}} \color[HTML]{F1F1F1} \color{white} {\cellcolor{white}}  & {\cellcolor[HTML]{000000}} \color[HTML]{F1F1F1} \color{white} {\cellcolor{white}}  & {\cellcolor[HTML]{000000}} \color[HTML]{F1F1F1} \color{white} {\cellcolor{white}}  & {\cellcolor[HTML]{000000}} \color[HTML]{F1F1F1} \color{white} {\cellcolor{white}}  \\
{\attIndent} Checks at run time & 0.14 & {\cellcolor[HTML]{000000}} \color[HTML]{F1F1F1} \color{white} {\cellcolor{white}}  & {\cellcolor[HTML]{C0E593}} \color[HTML]{000000} 0.50 & {\cellcolor[HTML]{000000}} \color[HTML]{F1F1F1} \color{white} {\cellcolor{white}}  & {\cellcolor[HTML]{000000}} \color[HTML]{F1F1F1} \color{white} {\cellcolor{white}}  & {\cellcolor[HTML]{000000}} \color[HTML]{F1F1F1} \color{white} {\cellcolor{white}}  & {\cellcolor[HTML]{000000}} \color[HTML]{F1F1F1} \color{white} {\cellcolor{white}}  & {\cellcolor[HTML]{000000}} \color[HTML]{F1F1F1} \color{white} {\cellcolor{white}}  & {\cellcolor[HTML]{000000}} \color[HTML]{F1F1F1} \color{white} {\cellcolor{white}}  \\
{\attIndent} Quality/security checks & 0.74 & {\cellcolor[HTML]{000000}} \color[HTML]{F1F1F1} \color{white} {\cellcolor{white}}  & {\cellcolor[HTML]{000000}} \color[HTML]{F1F1F1} \color{white} {\cellcolor{white}}  & {\cellcolor[HTML]{000000}} \color[HTML]{F1F1F1} \color{white} {\cellcolor{white}}  & {\cellcolor[HTML]{000000}} \color[HTML]{F1F1F1} \color{white} {\cellcolor{white}}  & {\cellcolor[HTML]{E89AC6}} \color[HTML]{000000} 0.25 & {\cellcolor[HTML]{000000}} \color[HTML]{F1F1F1} \color{white} {\cellcolor{white}}  & {\cellcolor[HTML]{000000}} \color[HTML]{F1F1F1} \color{white} {\cellcolor{white}}  & {\cellcolor[HTML]{000000}} \color[HTML]{F1F1F1} \color{white} {\cellcolor{white}}  \\
\attHeader{Composability} & & & & & & & & & \\
{\attIndent} Independent & 0.56 & {\cellcolor[HTML]{E283B7}} \color[HTML]{F1F1F1} 0.00 & {\cellcolor[HTML]{000000}} \color[HTML]{F1F1F1} \color{white} {\cellcolor{white}}  & {\cellcolor[HTML]{000000}} \color[HTML]{F1F1F1} \color{white} {\cellcolor{white}}  & {\cellcolor[HTML]{ACD977}} \color[HTML]{000000} 1.00 & {\cellcolor[HTML]{ACD977}} \color[HTML]{000000} 1.00 & {\cellcolor[HTML]{000000}} \color[HTML]{F1F1F1} \color{white} {\cellcolor{white}}  & {\cellcolor[HTML]{000000}} \color[HTML]{F1F1F1} \color{white} {\cellcolor{white}}  & {\cellcolor[HTML]{E283B7}} \color[HTML]{F1F1F1} 0.00 \\
{\attIndent} Vendor internal & 0.15 & {\cellcolor[HTML]{000000}} \color[HTML]{F1F1F1} \color{white} {\cellcolor{white}}  & {\cellcolor[HTML]{000000}} \color[HTML]{F1F1F1} \color{white} {\cellcolor{white}}  & {\cellcolor[HTML]{000000}} \color[HTML]{F1F1F1} \color{white} {\cellcolor{white}}  & {\cellcolor[HTML]{000000}} \color[HTML]{F1F1F1} \color{white} {\cellcolor{white}}  & {\cellcolor[HTML]{000000}} \color[HTML]{F1F1F1} \color{white} {\cellcolor{white}}  & {\cellcolor[HTML]{000000}} \color[HTML]{F1F1F1} \color{white} {\cellcolor{white}}  & {\cellcolor[HTML]{000000}} \color[HTML]{F1F1F1} \color{white} {\cellcolor{white}}  & {\cellcolor[HTML]{43861F}} \color[HTML]{F1F1F1} 1.00 \\
{\attIndent} Package manager type & 0.19 & {\cellcolor[HTML]{498D20}} \color[HTML]{F1F1F1} 1.00 & {\cellcolor[HTML]{000000}} \color[HTML]{F1F1F1} \color{white} {\cellcolor{white}}  & {\cellcolor[HTML]{000000}} \color[HTML]{F1F1F1} \color{white} {\cellcolor{white}}  & {\cellcolor[HTML]{000000}} \color[HTML]{F1F1F1} \color{white} {\cellcolor{white}}  & {\cellcolor[HTML]{000000}} \color[HTML]{F1F1F1} \color{white} {\cellcolor{white}}  & {\cellcolor[HTML]{000000}} \color[HTML]{F1F1F1} \color{white} {\cellcolor{white}}  & {\cellcolor[HTML]{000000}} \color[HTML]{F1F1F1} \color{white} {\cellcolor{white}}  & {\cellcolor[HTML]{000000}} \color[HTML]{F1F1F1} \color{white} {\cellcolor{white}}  \\
\attHeader{Analytics} & & & & & & & & & \\
{\attIndent} Sentiment and popularity ratings & 0.73 & {\cellcolor[HTML]{000000}} \color[HTML]{F1F1F1} \color{white} {\cellcolor{white}}  & {\cellcolor[HTML]{000000}} \color[HTML]{F1F1F1} \color{white} {\cellcolor{white}}  & {\cellcolor[HTML]{000000}} \color[HTML]{F1F1F1} \color{white} {\cellcolor{white}}  & {\cellcolor[HTML]{000000}} \color[HTML]{F1F1F1} \color{white} {\cellcolor{white}}  & {\cellcolor[HTML]{CD3A8D}} \color[HTML]{F1F1F1} 0.00 & {\cellcolor[HTML]{000000}} \color[HTML]{F1F1F1} \color{white} {\cellcolor{white}}  & {\cellcolor[HTML]{F1B5D9}} \color[HTML]{000000} 0.33 & {\cellcolor[HTML]{000000}} \color[HTML]{F1F1F1} \color{white} {\cellcolor{white}}  \\
{\attIndent} Marking feedback & 0.25 & {\cellcolor[HTML]{000000}} \color[HTML]{F1F1F1} \color{white} {\cellcolor{white}}  & {\cellcolor[HTML]{000000}} \color[HTML]{F1F1F1} \color{white} {\cellcolor{white}}  & {\cellcolor[HTML]{000000}} \color[HTML]{F1F1F1} \color{white} {\cellcolor{white}}  & {\cellcolor[HTML]{000000}} \color[HTML]{F1F1F1} \color{white} {\cellcolor{white}}  & {\cellcolor[HTML]{000000}} \color[HTML]{F1F1F1} \color{white} {\cellcolor{white}}  & {\cellcolor[HTML]{000000}} \color[HTML]{F1F1F1} \color{white} {\cellcolor{white}}  & {\cellcolor[HTML]{000000}} \color[HTML]{F1F1F1} \color{white} {\cellcolor{white}}  & {\cellcolor[HTML]{000000}} \color[HTML]{F1F1F1} \color{white} {\cellcolor{white}}  \\
{\attIndent} Product Usage data & 0.33 & {\cellcolor[HTML]{000000}} \color[HTML]{F1F1F1} \color{white} {\cellcolor{white}}  & {\cellcolor[HTML]{000000}} \color[HTML]{F1F1F1} \color{white} {\cellcolor{white}}  & {\cellcolor[HTML]{000000}} \color[HTML]{F1F1F1} \color{white} {\cellcolor{white}}  & {\cellcolor[HTML]{000000}} \color[HTML]{F1F1F1} \color{white} {\cellcolor{white}}  & {\cellcolor[HTML]{000000}} \color[HTML]{F1F1F1} \color{white} {\cellcolor{white}}  & {\cellcolor[HTML]{000000}} \color[HTML]{F1F1F1} \color{white} {\cellcolor{white}}  & {\cellcolor[HTML]{000000}} \color[HTML]{F1F1F1} \color{white} {\cellcolor{white}}  & {\cellcolor[HTML]{000000}} \color[HTML]{F1F1F1} \color{white} {\cellcolor{white}}  \\
\attHeader{Communication channels} & & & & & & & & & \\
{\attIndent} Documentation (wikis, FAQs) & 0.81 & {\cellcolor[HTML]{000000}} \color[HTML]{F1F1F1} \color{white} {\cellcolor{white}}  & {\cellcolor[HTML]{000000}} \color[HTML]{F1F1F1} \color{white} {\cellcolor{white}}  & {\cellcolor[HTML]{000000}} \color[HTML]{F1F1F1} \color{white} {\cellcolor{white}}  & {\cellcolor[HTML]{000000}} \color[HTML]{F1F1F1} \color{white} {\cellcolor{white}}  & {\cellcolor[HTML]{E283B7}} \color[HTML]{F1F1F1} 0.25 & {\cellcolor[HTML]{000000}} \color[HTML]{F1F1F1} \color{white} {\cellcolor{white}}  & {\cellcolor[HTML]{000000}} \color[HTML]{F1F1F1} \color{white} {\cellcolor{white}}  & {\cellcolor[HTML]{000000}} \color[HTML]{F1F1F1} \color{white} {\cellcolor{white}}  \\
{\attIndent} Product homepage & 0.97 & {\cellcolor[HTML]{000000}} \color[HTML]{F1F1F1} \color{white} {\cellcolor{white}}  & {\cellcolor[HTML]{000000}} \color[HTML]{F1F1F1} \color{white} {\cellcolor{white}}  & {\cellcolor[HTML]{000000}} \color[HTML]{F1F1F1} \color{white} {\cellcolor{white}}  & {\cellcolor[HTML]{000000}} \color[HTML]{F1F1F1} \color{white} {\cellcolor{white}}  & {\cellcolor[HTML]{000000}} \color[HTML]{F1F1F1} \color{white} {\cellcolor{white}}  & {\cellcolor[HTML]{000000}} \color[HTML]{F1F1F1} \color{white} {\cellcolor{white}}  & {\cellcolor[HTML]{000000}} \color[HTML]{F1F1F1} \color{white} {\cellcolor{white}}  & {\cellcolor[HTML]{000000}} \color[HTML]{F1F1F1} \color{white} {\cellcolor{white}}  \\
{\attIndent} Star/Score/Up/Downvote rating & 0.57 & {\cellcolor[HTML]{EBA1CB}} \color[HTML]{000000} 0.11 & {\cellcolor[HTML]{B0DC7D}} \color[HTML]{000000} 1.00 & {\cellcolor[HTML]{000000}} \color[HTML]{F1F1F1} \color{white} {\cellcolor{white}}  & {\cellcolor[HTML]{B0DC7D}} \color[HTML]{000000} 1.00 & {\cellcolor[HTML]{E07EB3}} \color[HTML]{F1F1F1} 0.00 & {\cellcolor[HTML]{B0DC7D}} \color[HTML]{000000} 1.00 & {\cellcolor[HTML]{E07EB3}} \color[HTML]{F1F1F1} 0.00 & {\cellcolor[HTML]{000000}} \color[HTML]{F1F1F1} \color{white} {\cellcolor{white}}  \\
{\attIndent} Written reviews (in text) & 0.47 & {\cellcolor[HTML]{EBA1CB}} \color[HTML]{000000} 0.00 & {\cellcolor[HTML]{000000}} \color[HTML]{F1F1F1} \color{white} {\cellcolor{white}}  & {\cellcolor[HTML]{000000}} \color[HTML]{F1F1F1} \color{white} {\cellcolor{white}}  & {\cellcolor[HTML]{91C857}} \color[HTML]{000000} 1.00 & {\cellcolor[HTML]{EBA1CB}} \color[HTML]{000000} 0.00 & {\cellcolor[HTML]{B2DD7F}} \color[HTML]{000000} 0.89 & {\cellcolor[HTML]{EBA1CB}} \color[HTML]{000000} 0.00 & {\cellcolor[HTML]{000000}} \color[HTML]{F1F1F1} \color{white} {\cellcolor{white}}  \\
{\attIndent} Community Forum & 0.45 & {\cellcolor[HTML]{000000}} \color[HTML]{F1F1F1} \color{white} {\cellcolor{white}}  & {\cellcolor[HTML]{000000}} \color[HTML]{F1F1F1} \color{white} {\cellcolor{white}}  & {\cellcolor[HTML]{CFEBAA}} \color[HTML]{000000} 0.75 & {\cellcolor[HTML]{000000}} \color[HTML]{F1F1F1} \color{white} {\cellcolor{white}}  & {\cellcolor[HTML]{ECA6CF}} \color[HTML]{000000} 0.00 & {\cellcolor[HTML]{000000}} \color[HTML]{F1F1F1} \color{white} {\cellcolor{white}}  & {\cellcolor[HTML]{000000}} \color[HTML]{F1F1F1} \color{white} {\cellcolor{white}}  & {\cellcolor[HTML]{000000}} \color[HTML]{F1F1F1} \color{white} {\cellcolor{white}}  \\
{\attIndent} Support Ticket & 0.35 & {\cellcolor[HTML]{000000}} \color[HTML]{F1F1F1} \color{white} {\cellcolor{white}}  & {\cellcolor[HTML]{000000}} \color[HTML]{F1F1F1} \color{white} {\cellcolor{white}}  & {\cellcolor[HTML]{000000}} \color[HTML]{F1F1F1} \color{white} {\cellcolor{white}}  & {\cellcolor[HTML]{000000}} \color[HTML]{F1F1F1} \color{white} {\cellcolor{white}}  & {\cellcolor[HTML]{000000}} \color[HTML]{F1F1F1} \color{white} {\cellcolor{white}}  & {\cellcolor[HTML]{000000}} \color[HTML]{F1F1F1} \color{white} {\cellcolor{white}}  & {\cellcolor[HTML]{000000}} \color[HTML]{F1F1F1} \color{white} {\cellcolor{white}}  & {\cellcolor[HTML]{000000}} \color[HTML]{F1F1F1} \color{white} {\cellcolor{white}}  \\
{\attIndent} Promotion/Marketing & 0.71 & {\cellcolor[HTML]{000000}} \color[HTML]{F1F1F1} \color{white} {\cellcolor{white}}  & {\cellcolor[HTML]{000000}} \color[HTML]{F1F1F1} \color{white} {\cellcolor{white}}  & {\cellcolor[HTML]{000000}} \color[HTML]{F1F1F1} \color{white} {\cellcolor{white}}  & {\cellcolor[HTML]{000000}} \color[HTML]{F1F1F1} \color{white} {\cellcolor{white}}  & {\cellcolor[HTML]{EBA3CD}} \color[HTML]{000000} 0.25 & {\cellcolor[HTML]{000000}} \color[HTML]{F1F1F1} \color{white} {\cellcolor{white}}  & {\cellcolor[HTML]{000000}} \color[HTML]{F1F1F1} \color{white} {\cellcolor{white}}  & {\cellcolor[HTML]{000000}} \color[HTML]{F1F1F1} \color{white} {\cellcolor{white}}  \\
\bottomrule
\end{tabular}
\end{adjustbox}
\end{table*}
 
The table only shows the top 10 deviations per cluster (i.e., \emph{column}) to focus on the most important contributors to each cluster.
Since all \tbdWdims are binary --- each store has or does not have the \tbdWdim --- all values of the centroid-of-centroids are between $[0,1]$; thus, a positive deviation (shown with a green background) implies that the stores in the cluster are \emph{more} likely to have the attribute, and a negative deviation (shown with a magenta background) implies that the stores are \emph{less} likely to have the attribute.

For example, for cluster 8 the most important contributor is \emph{[Composability] Vendor internal add-on/extension/unlock} where the centroid of the cluster is $1$. When comparing against the centroid-of-centroids (at $0.15$), the deviation is at $0.85$; this implies that all stores in this cluster have this \tbdWdim. 
On the other hand, an example of negative deviation for cluster 1 is the \tbdWdim \emph{[Composability] Independent} with a centroid of $0$ indicating that no stores in this cluster have this \tbdWdim.
Since the centroid-of-centroids for this \tbdWdims is at $0.56$, this implies the deviation for stores in this cluster is $-0.56$.

After the top characteristics that make each cluster distinctive had been identified, we leveraged this information to name and describe each cluster accordingly.
Using the information from \Cref{table:centroid_deviations} which shows the defining features of each cluster, we derived an organization of the clusters based on several dimensions.
The results are described in \Cref{table:cluster_des}.

\begin{sidewaystable*}
\scriptsize
\setlength\tabcolsep{4pt}
\newcommand{\wzIndentWidth}{2mm}
\setlength\extrarowheight{1pt}
\centering
\caption{List of stores and descriptions by cluster, with the example store
    that is closest to cluster centroid}
\label{table:cluster_des}
\begin{tabular}{@{}lp{35mm}>{\raggedright}p{35mm}p{70mm}r@{}}
\toprule
Type & Stores in Cluster & Example Store & Cluster Description & Index\\
\midrule
{\bf Extensions} & & & & \\
\hspace{\wzIndentWidth} Commercial specialized  & Adobe Magento, AutoDesk, BigCommerce, GoG, HubSpot, Plugin Boutique, Presta Shop, SketchUcation, CS-Cart & \emph{Presta Shop} offers addons to the ecommerce solution platform.                                                                         & Products in the stores are very domain specific. Creators are mostly business and their store front offers rating systems and written reviews.                                                                                &  6\\
\hspace{\wzIndentWidth} Community specialized & Bukkit, CurseForge, DockerHub, Home Assistant, Jmeter, Kodi, Minion, VSCode Marketplace                              & \emph{Kodi} add-on components offers extensions to the \emph{Kodi} entertainment center.                                                     & These are community focused stores that offers free products to users. Stores also tailor to a specific domain.                                                                                                               &  3\\
\hspace{\wzIndentWidth} Community non-specialized     & Apkpure, Chrome Web Store, Eclipse Marketplace, Firefox Add-ons, Gnome, Wordpress                                & \emph{Wordpress} offers free extensions for users using the wordpress platform.                                                              & Products in these stores offers extensions to the platform. Essential operations do not need registration (e.g., installing apps). Products offered in the stores face a generic audience and are independent from each other.&  4\\
{\bf General} & & & &\\
\hspace{\wzIndentWidth} Commercial  & AWS, Google Play Store, Microsoft Store, Nintendo eShop, Steam, Samsung Galaxy Store, Apple App Store     & \emph{MicroSoft Store} offers applications for the windows platform.                                                                         & Typical stores many people encounter everyday. They run for profit and offer vendor internal products supporting most monetization options.                                                                      &  8\\
\hspace{\wzIndentWidth} Community  & Chocolatey, F-Droid, Flatpak, Guardian
    Project Repository, IzzyOnDroid, Snapcraft                       &
    \emph{F-Droid} is a free and open source software only \android application store.                                                     & These stores contain standalone free products only. Creators for the stores are mostly from the community and the products are majority open source.                                                                          &  7\\
{\bf Package Manager}                   & Ajour, Jenkins, MacPorts, NPM, NuGet, Packagist, PyPI, Typo3, Ubuntu packages                            & \emph{Packagist} is the main repository for \emph{PHP} packages.                                                                             & No account system is involved for these stores. Products are free and most in package style with inter-dependency relationships. Communication channels are also limited with ratings and reviews missing for most stores.    &  1\\
{\bf Subscription oriented}    & Github Marketplace, JetBrains, Qt Marketplace, concrete5 marketplace                                     & \emph{Github Marketplace} offers applications and actions to improve the workflow related to \emph{git} repositories hosted on \emph{GitHub} & Often offers subscription services and supports DRM management by the store. Products are not standalone applications and either provide service or extends a platform.                                                       &  2\\
{\bf Other}     & MarketPlaceKit, RICOH THETA, Sellacious, daz3D                                                           & \emph{Sellacious} is a ecommerce platform and provides extensions to the platform.                                                           & They do not have much communication channels offered. Rating and reviews do not exist in the stores. The stores mostly exists to distribute extensions centrally to the platform they are based on.                            &  5\\
\bottomrule
\end{tabular}
\end{sidewaystable*}
 
One important dimension focuses on the type of application served by stores in the cluster.
We identified three major types of applications that differentiate the clusters: \emph{General}, where the store offers stand-alone programs that run without the need of specific software (aside from a specific operating system, e.g., \googlePlay, \name{AWS}, \name{Steam}); \emph{Extensions}, where the store offers extensions to a specific program or platform e.g., \name{VSCode Marketplace} for \emph{VSCode}, \name{Chrome Web Store} for \emph{Google Chrome};  and \emph{Package manager}, where the store offers stand-alone programs, but also manages dependency-relationships and requirements between different applications in the store e.g., \name{NPM}, \name{MacPorts}, \name{Ubuntu Packages}.
Another dimension in which these clusters can be organized is whether they are Commercial (business-oriented) or Community-managed (no money is involved).

\begin{hassanbox}{\columnwidth}[4ex]
    App stores are not all alike.
    Intuitive groupings emerge naturally from the data.
    Their differences can be due to the type of application they offer --- standalone or extensions --- and their operational model, either business- or community-oriented.
    We found that app stores in different groups of our clustering have different properties, and these properties may have bearing on empirical studies involving app stores.
\end{hassanbox}

\section{Discussion}
\label{sec:discussion}

In this section, we discuss our findings regarding what we consider \appStores to be based on our clustering results,
and we describe various research opportunities involving the influence of \appStores on software engineering practices.

\subsection{What Is an App Store?}

The term \emph{app store} became popular largely through \appleAppStore, which launched in 2008 along with the
\emph{iPhone 3G}~\cite{app.store.introduce}.
Other online software stores have also appeared and have had the term applied to them.
Originally, the term usually referred to stores of applications for mobile devices, but we have found that today there is ample diversity of the type of applications that \appStores offer and in the features they provide to app developers and users.
\AppStores are also dynamic: features are continually being added, removed, and altered by store owners in response to changes in their goals and feedback from their socio-technical environments.
For example, the \name{Chrome Web Store} initially introduced a built-in monetization option that provided a mechanism for applications to receive payments from its users; however, the store later decided to deprecate this monetization option~\cite{chrome.disable.payment} and suggested developers to switch to alternative payment-handling options.

In our work, we have employed a working definition through our inclusion/exclusion criteria for app stores to be included in our research.
However, due to the complexity, diversity, and constantly evolving nature of app stores, we have decided not to attempt a firm, prescriptive definition of the term.
Instead, in the following paragraphs, we will discuss each of several aspects of \appStores in detail, and hope that in the future, a more robust definition and operating model can emerge.

\subsubsection{Common Features of App Stores}

Although we found significant diversity among the example \appStores we studied, we were able to identify a set of three common features that appear to span the space of \appStores.

\paragraph{Simple installation and updates of apps}

An \appStore facilitates simple installation of a selected application, and can also enable simple updating.
For some stores, apps are expected to run on the hardware of the client; in others, the \appStore provides and manages the hardware where the app runs.
In both cases, the \appStore frees the user from worrying about the technical details of installation, including compatibility with their specific hardware and software configuration, as well as the installation of the app and its dependencies, if any. 
Typically, app stores will also automate the installation of updates to the application, again freeing the user from worrying about if they have the latest version of the app with the latest features and bug fixes.

\paragraph{App exploration and discovery}

\AppStores provide mechanisms that allow users to find apps they might want to use.  
In its simple form, this mechanism might be a search engine that returns a list of apps that match a given set of keywords (such as \name{homebrew}, \name{PyPI}). 
In \tbdWlabelledset, 73\% of stores provide some kind of aggregated recommendations (e.g., advertisement and trends in \name{WordPress}), up to personal recommendations that are based on other apps the user has installed before (e.g., \appleAppStore). 
User feedback via reviews (present in 47\% of \tbdWlabelledset) and forums (present in 45\% of \tbdWlabelledset) can provide further information to aid other users in identifying apps of possible interest to them.

\begin{figure}
    \includegraphics[width=\columnwidth]{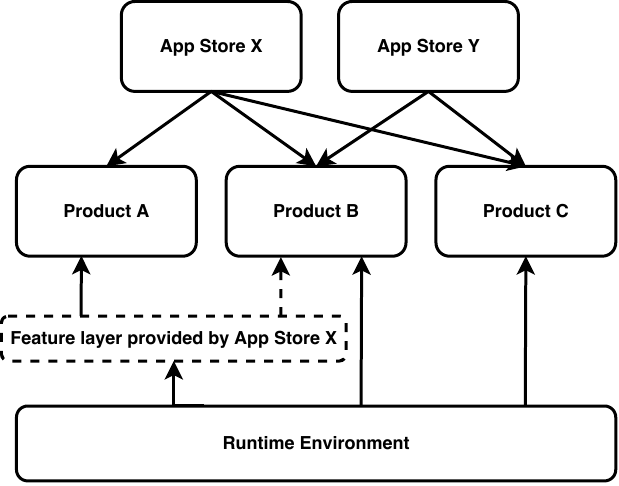}
    \caption{Stores may offer optional extensions to the runtime environment for applications}
\label{fig:store_rte}
\end{figure}

\paragraph{The \appStore guarantees the runtime environment}
In practice, app stores often execute within a runtime environment (RTE),
such as an operating system (e.g., \googlePlay on \android) or an extensible software application (e.g., \name{Firefox Add-ons} on \emph{Firefox}).
Many app stores simply sit on top of the RTE, acting primarily as a gatekeeper for adding and deleting apps.  
However, some app stores are more tightly integrated with the RTE; in extreme cases, the app store can extend the RTE with the app store's own functionality and together provide an augmented RTE for the applications managed through the app store.
\name{Steam} is a good example for extending the RTE with its own features; developers can integrate with many services offered by \name{Steam}, such as an achievement system that offers players recognition when they fulfill certain requirements in the game.
\Cref{fig:store_rte} illustrates the situation where a product may integrate with additional store-added features to the RTE, which in turn enriches the user experience of the store users.
When \emph{Product B} is offered in \emph{App Store Y}, it will not have the features provided by \emph{App Store X}.

The \appStore ensures that apps are installed only when their runtime requirements are satisfied.
The process is often done through running checks on apps submitted to the app store, which 74\% of \tbdWlabelledset perform specifically.
By specifying the runtime requirements, the assumption for both the developer and the user is that if the application is installed --- implying that the requirements are satisfied --- it is expected to run properly.
This is usually achieved by a software layer on top of the RTE, provided by either the \appStore or the user. 
In its simplest form, this software layer is responsible for installing and updating apps (see ``Simple installation and updates of apps'' above).
In some cases, this software layer might also include a set of libraries that the apps can use to provide features specific to the \appStore thus forming part of the RTE for the applications. 
These libraries might range in purpose (domain specific, common GUI, resource management, etc.). 
In extreme cases, this layer includes the operating system, as it is the case with \appleAppStore.  
However, checks during runtime is a very rare feature, which on 14\% of \tbdWlabelledset provides.

Some hardware platforms have become so tightly integrated to the software layer of the \appStore that they can be considered monolithic: the hardware is rendered unusable without the \appStore. 
This is exemplified by the \appleAppStore, where one cannot use the hardware without first having an account in the \appStore; even operating system upgrades are distributed via the store.

This tight level of integration has clear benefits for all three stakeholders: end users have fewer installation technical details to worry about; app developers can be assured that users will be able to install their apps without the need for technical support; and app store owners can strictly manage who has access to the user's RTE and how.
However, such tight integration is technically unnecessary and may even be undesirable.
From a software engineering perspective, such tight coupling could be seen as a ``design smell'', since the operating system and the app store layers address fundamentally different concerns.  
Also, tight integration can create an artificial barrier to competition, effectively establishing a quasi-monopoly for the store owner; the store owner may assume the role of gatekeeper not only for streamlining technical issues, but also for business reasons, requiring a kind of toll to be paid by app developers for access to the store.
A recent initiative in the European Union~\cite{eu.commission} aims to enable fair competition by enforcing that ecosystems are opened up, which will likely also allow the installation of alternative software layers for other \appStores, a term called side-loading.
In contrast to the \emph{Apple}'s tight control of the operating system as
part of its \appStore, \android allows third-party \appStore software (e.g., \name{F-Droid}~\cite{store.fdroid}) to be installed in co-existence with the system default (often \googlePlay).

As mentioned above, some stores distribute software that runs on hardware owned by the \AppStore itself; in these cases, the RTE is fully managed and controlled by the store.
For example \name{GitHub Marketplace} and \name{Atlassian Marketplace} offer applications that run on GitHub and Atlassian servers respectively.
In most cases, these applications are not deployed to the user's computers.

\subsubsection{Different Types of App Stores}

While some features are broadly shared by all app stores,
in \Cref{sec:data-analysis}, we identified different groups of app stores based on their features.
For stores within the same group, they often share common features, whereas across different groups, the stores tend to have less in common.
We now discuss the differences across the groups in detail.

\paragraph{Diversity in goals}

As a platform focusing on delivering products to customers, the high-level goal of one app store can be dramatically different from the other.
Even \appStores providing software for the same underlying RTE can have radically different purposes.
For example, consider the \appStores that run on \android.
\googlePlay is the \emph{de facto} store for \android applications.
\name{F-Droid} store, on the other hand, offers only free and open source \android applications, and \name{APKPure} offers multiple versions of the same software so the user can decide which version they would like to install.

\emph{Apple's} \appStore offers applications for all its RTEs: \emph{MacOS} (laptop and desktops), \emph{iOS} (phones and tables), and the \emph{Safari} browser.
In contrast, \emph{Google} has different stores for \emph{AndroidOS} and for its web browser, \emph{Chrome}.
The \name{Microsoft Store} sells hardware and apps for \emph{Windows} and \emph{XBox}.
\name{Alexa Skills} offers skills that enhance the voice agent \emph{Alexa}'s capabilities.

In many program language ecosystems, the core language development (focusing on the language features) and packaging system (focusing on extending the functionality of the language) are led by separate organizations (e.g., NPM~\cite{org.npm} and JavaScript~\cite{org.javascript}).

\paragraph{Diversity in business model}

Another important difference we observed is between business-managed and community-managed stores.
In business-managed stores (with few exceptions), a primary goal is to generate a profit.
These stores provide a payment mechanism between the app creator and the purchaser, with the store keeping a percentage of any sales.  
These stores have to solve three key concerns: first, implementing registration and authentication of users and developers; second, some type of digital rights management, so only users who have acquired the software can use it; and third, a payment mechanism e.g., subscription, one-time payment, and advertisement.  

Community-managed stores, on the other hand, are often run by volunteers, and their features focus on facilitating not-for-profit product delivery from developer to user.
Many community stores offer limited community interactions compared to business stores where customer feedback is important.
For example, in the \name{Kodi} store, add-ons have a web page (e.g., \emph{The Movie Database Python}~\cite{kodi.tmdb}).
This page provides information regarding installation of the add-on, such as known compatibility concerns, download links, and installation requirements.
Meanwhile, most communication channels about the add-on are hosted elsewhere; for example, installation and usage instructions, extended descriptions, and screenshots can be found in the community forum instead.

It is important to note that the products contained in community-oriented stores are not limited to open source software; some community-managed \appStore policies often permit the distribution of proprietary software.
In the natural groupings we observed, no rights management are enforced from the store side for Cluster 3, meanwhile, most stores in Cluster 8 have some form of rights management built-in to the store.
For example, \name{Homebrew} permits apps that are not open source if the apps are free to use; these apps might include in-app purchases --- such as an upgrade to a full-feature app --- that are handled outside of \name{Homebrew}.

\subsection{Implications for the Main Participant Stakeholders}

The results of our study includes an evidence-based detailed view of the broad landscape of app stores.
This view can help us improve the understanding of the realities and potentialities of app stores in general.
Meanwhile, the results of our work can also benefit the different stakeholders involved with \appStores, including app creators, app stores themselves, users, and researchers.

\paragraph{Application creators}
Those who create applications --- including those who design, develop, test, and market apps --- benefit from a holistic view of other stores that will allow them identify potential new markets (stores where they can offer their software) and to understand changing and emerging features that could eventually come to their app store of choice. 
For new creators, this research emphasizes that a software store has both technical requirements --- such as the use of a specific software development kit --- and non-technical ones --- such as restrictions on what applications can do, approval processes and timelines --- and that these requirements vary significantly from one store to another.

\paragraph{App-stores}
The overview presented herein provides a framework for comparison between app-stores, particularly those that operate on the same market, such as \android application stores. 
It can also help promote wide adoption of features that are not universal, such as communication channels between users and developers.

\paragraph{Users}
With the diversity in app stores, especially when multiple app stores are competing in the same domain, it allows users the chose where to acquire their applications.
This allows for more diversity for how the apps are distributed and the user's choice also affect the competition.

\paragraph{Researchers}
As discussed in \Cref{sec:background}, most prior research has focused on the applications offered in \appStores, and there is a need for research that focuses on studying the store themselves.
This emphasis could aid researchers in considering different points of view when conducting app store-centric studies, and also suggest avenues of exploration concerning how the development process is affected by the existence of app stores.

We describe this point in detail in the next sections.

\subsection{App Store Features}

In this section, we discuss how each of feature groups from \Cref{tab:dim_listing} has been addressed by current SE research and we suggest some possible future directions.

\paragraph{Monetization}
App development can be affected by their pricing strategy.
For example different software architecture to support a different system of monetization (e.g., locking functionality behind microtransactions)~\cite{pham2017paas}.
Studies have shown a correlation between app features and pricing~\cite{finkelstein2017investigating,harman2012app,sarro2015feature}.
Moreover, in many studies on apps~\cite{alshayban2020accessibility,aljedaani2019comparison}, free apps and charged apps are often considered as different types of applications.
Future work could further explore how different monetization options affect app development.

\paragraph{Rights management}
Digital rights management is still an ongoing challenge in software engineering.
Existing studies have explored the options of implementing different DRM systems to support developers~\cite{lu2019blockchain,gaber2020privdrm}.
DRM can also add challenges in other development activities such as complicating the testing procedure~\cite{sung2019test} and affect performance~\cite{denuvo.performance}.
Often we can observe the store offering means for providing and enforcing DRM.
Because DRM is still a nascent technology within software engineering, it remains an open area to explore for future study and how app stores can play a role.

\paragraph{Account requirement}
User identity enables telemetry of user behavior.
An account system is also the prerequisite of a store-wide DRM system as discussed in the previous paragraph.
Existing research has focused on how to leverage the user identity information to create targeted recommender systems~\cite{he2015mining} and also investigated the concerns of privacy-related issues~\cite{scoccia2022empirical}.
The interest of developers (detailed tracing data) and users (privacy) are in conflict, app stores that require user identification could prove to be an excellent study subject for future research in that area.

\paragraph{Product type}
Existing research has already shown different software engineering practices based on the software product.
For example, gaming development is very different from traditional software development and open source development~\cite{murphy2014cowboys,pascarella2018video}.
Research have shown that different types of software can introduce specific challenges unique to them~\cite{lee2020empirical,ibrahim2020too}.
Future research should better understand how the product type affects user expectations and development practices, for example, with respect to the delivery of software or the way creators and users can interact.

\paragraph{Target audience}
When an app developer decides on a specific app store to sell their app, they are also effectively selecting for a specific type of user~\cite{subramanian2006empirical,manotas2016empirical,gholami2021should}.
Users of a general-purpose store such as \googlePlay are different and much more diverse than the user population~\cite{guzman2018user} in very specialized stores, such as the add-on store for a particular game.
Research needs to understand better which features are relevant in each specific context~\cite{sun2021empirical}, so the experience can be tailored to the concrete situation.

\paragraph{Type of product creators}
Existing research has shown many differences between open source and industrial software development~\cite{pascarella2018video,lee2020building}.
Some studies have touched the aspect of release engineering in open source development~\cite{nayebi2017version}, where developers would strategically select which versions to release on the app store.
However, we believe that there is still room for more understanding in how targeting releases towards app stores affects software development.

\paragraph{Intent of app store}
While in most domains, there exists a dominant app store, we can also observe situations where multiple app stores compete in the same domain (e.g., game stores on PC, mobile app stores in China~\cite{wang2018beyond}).
In these situations, users have a choice of which app store to use when the same application is offered.
In practice, some studies have explored how the high level operation of app stores can affect the software delivery process especially involving security concerns~\cite{hu2021champ,sun2021empirical}.
Competition between app stores within the same domain remains largely unstudied, as does how their operations can affect both developers and users.

\paragraph{Role of intermediary}
App stores provide a platform for users and developers.
Researchers have explored how it affects software development processes such as testing and release management~\cite{nayebi2016release,shen2017towards}.
There are many opportunities for security~\cite{ferreira2021containing} and quality assurance~\cite{al2019app,tang2019large} to be ensured on the app store side.
Future study can explore how the differences between apps managed through an app store and apps that are not.
For example, studying the difference between open-source web extensions that are in and not in app stores.

\paragraph{Composability}
Existing research has explored co-installability in the scope of package manager systems~\cite{vouillon2013software,claes2015historical}.
However, we only have limited understanding of co-installability for standalone applications in an extension system.
For example, if two standalone extensions were to modify the same component of the underlying software, a potential incompatibility could occur.
Future research can explore this area by performing empirical studies on existing systems to understand the issue of conflicts.

\paragraph{Analytics}
App stores as the central hub between developers and users have access to rich information useful for analytics.
Previous studies have taken advantage of the app store specific information to help software developers~\cite{mcmillan2012detecting,martin2016causal,maalej2019data}.
For example, Ullmann~\etal{}~\cite{ullmann2022makes} leveraged records of rating statistics and downloaded information to study the factors in developing successful video games.
Another study leveraged analytic information collected by the app store to identify incompatible builds of application and physical devices~\cite{khalid2014prioritizing}.
Future work can explore what are the possible data to collect and form analytics, and how can the analytic data be leveraged to help developers and users.

\paragraph{Communication channels}
Communication channels are the most studied area of app store features.
Specifically, there has been a heavy focus on app reviews, where researchers have leveraged the information in app reviews to aid software development in areas such as extracting/locating bug reports~\cite{haering2021automatically}, discover feature requests~\cite{wu2021identifying} and collect user feedback~\cite{guo2020caspar}.
However, existing studies also suggest that the use of communication channels in app stores are often multi-purpose~\cite{dkabrowski2022analysing}.
Researchers also find that some interaction requirements between interested parties are relegated to other platforms such as Twitter~\cite{nayebi2017app}.
Future work can explore different types of communication channels in their functionality and how they can integrate with app stores.
The corpora from communication channels are also rich information sources where researchers can leverage to extract information about developer-user interactions.

\subsection{Research Opportunities Involving \AppStores}

\AppStores are becoming the primary channel for software delivery and exert considerable influence in many aspects of the software development process.
A previous study by Rosen and Shihab~\cite{rosen2016mobile} on Stack Overflow questions by mobile developers has shown that app delivery is one of the biggest challenges developers faced.
Our results in \Cref{sec:data-analysis} demonstrate that there is a wide variety of types of stores, each with different features and goals.
Today, \appStores encompass many kinds of applications, from games running on the hardware of the user to add-ons for applications that run on corporate servers such as \emph{GitHub}.
However, existing research often focuses heavily on the applications offered inside \appStores, especially those of the two major mobile app stores.
In the following paragraphs, we discuss several research opportunities to study how app stores can affect software development.

\subsubsection{\AppStores as Actors in Software Development}

\paragraph{\AppStores affect the software product cycle}

Researchers need to consider how and why \appStores can affect the software development life cycle.
For example, we know that \appStores can constrain and sometimes even dictate software release processes.
Some stores go beyond this and exert a kind of socio-technical environmental pressure on other software development practices, becoming a de facto stakeholder in app development.
Sometimes these environmental pressures are technical in nature, where the app store might dictate the programming language or deployment platform/OS; some \appStores go further and create RTEs, software development kits (SDKs), and user interface (UI) libraries that must be used by all app developers.
Sometimes these environmental pressures are non-technical in nature, such as when the app store prescribes the kinds of application that is allowed in the store.
For example, \emph{Microsoft} recently announced that it will not permit app developers to profit from open source applications.\footnote{See Update to 10.8.7 \url{https://docs.microsoft.com/en-us/windows/uwp/publish/store-policies-change-history}}
When an app store operates in a manner such that it has control over what kind of application to include,
it creates a software ecosystem and as such, it faces the same challenges that any other ecosystem has: how to thrive.
In particular, stores need to understand the needs of their developers and users to retain existing ones and attract new ones.
However, suggested by what we have observed in \Cref{sec:rq1}, \appStores are diverse with a large number of features that characterize and differentiate between them.
While stores are experimenting and evolving, each action is likely to have an effect on the ecosystems they formed, both positively and negatively.
Thus, the impact of \appStores in the economy and their markets is worthy of further study.

\paragraph{An app may be offered in several app stores}

Developers want to run their software on the platform that is provided or supported by the store, and as such they must accept the requirements and limitations that such a store may impose.
This issue is compounded when the app is being offered in more than one store, as the developers might have to adapt their processes to different sets of requirements, some of which might be conflicting.
For instance, an app can be both available in \name{F-Droid} (in Cluster 7) and \googlePlay (in Cluster 8).
In \googlePlay, it is common for applications to collect telemetry data to better understand typical user behaviour; however, in \name{F-Droid} --- an open source and privacy-oriented store --- such data collection is highly discouraged.
Furthermore, developers must also adapt to the features and limitations that a store provides regarding software deployment, communication with users and --- when they exist --- the mechanism available to profit from their software and to use digital rights management.
This is particularly interesting if the targeted app stores are in different natural groupings.
This introduces new areas of studies such as how store policies propagate to applications over time, and how violations of store policies can be detected automatically.
Researchers have already begun to investigate this topic through qualitative approaches to identify how applications comply with specific policies that concern accessibility~\cite{alshayban2020accessibility} and human values~\cite{obie2021first}.

\paragraph{App stores strongly affect the release engineering process}
\AppStores are especially important in release engineering.
Specifically, the release process needs to consider how the application is to be packaged, deployed, and updated.
The heterogeneity of the platform provided by RTEs might also affect the number of versions of the application that need to be deployed, e.g., variety of target CPUs, different screen sizes and orientations, and amount of available memory.

When an application is developed for multiple stores, it must effectively be managed as a product line; this is because multiple deliverables must be created, one for each platform-store combination~\cite{wang2019characterizing}.
Multiple deliverables can also help for telemetry reasons such as tracking the installation source of the application~\cite{ng2014android}.
The differences between packaged versions might be as significant as requiring the source code to be written in different programming languages, using different frameworks; also, each store is likely to require different deployment processes.

For example, when cross-releasing browser add-ons, developers may have to rewrite part of the functionality in \emph{Swift}/\emph{Objective-C} for better integration with Safari (in the \appleAppStore), while at the same time maintaining a fully \emph{JavaScript} version for \name{Chrome Web Store}.
Also, the scheduling of release activities is often dictated by the release processes of the stores.
A previous study has showed that taking into consideration of app review times is an important factor when planning releases~\cite{al-subaihin2021app}.
The \appStore standardizes, and often simplifies, the release engineering processes for its store; but it also becomes a potential roadblock that might delay or even reject a new release.

\subsubsection{The Challenge of Transferring Understanding Between Stores}

As noted above, prior work has examined many aspects of app stores, yet the app store itself has rarely been the focus of the research.
In many studies, the \appStore serves as a convenient collection of apps, and the research focuses on mobile development concerns such as testing and bug localization.
Even when research focuses on the app store itself, the scope rarely extends beyond \googlePlay and \appleAppStore.
Based on our observations, the diversity of app stores in their operational goals, business models, delivery channels, and feature sets can affect the generalizability of research outcomes.
For example, there have recently been many studies~\cite{dkabrowski2022analysing,lin2019empirical,wu2021identifying,haering2021automatically,guo2020caspar,obie2021first,fischer2021does} that focus on app reviews.
However, for an app store that does not have reviews (e.g., \name{Nintendo eShop}) none of the findings and tools can be leveraged (e.g., stores in Cluster 1, 5, and 7).

\paragraph{\AppStores that have the same features may still differ significantly}
Depending on the problem domain, the details of software development practices can vary dramatically.
For example, game development has been compared to both more traditional industrial software development~\cite{murphy2014cowboys} and to open source software development~\cite{pascarella2018video}; in both cases, the development processes can differ greatly.
We conjecture that the same may also occur across app stores, where despite the same feature is being offered in the different stores, the convention of using them could be different.
As mentioned above, one specific observation has been made between the gaming-focused store \name{Steam} and mobile stores (e.g., \googlePlay) in Cluster 8, where Lin~\etal{}~\cite{lin2019empirical} found that reviews across the platforms for the same app were often quite different in tone.
Such uncertainly invites future research to validate their findings in one store to another to improve the generalizability of the results, and also encourages replication studies to verify existing results on other stores.

\paragraph{A feature not in the \appStore does not mean the functionality is missing}
While some \appStores aim to provide a complete experience, where all interactions from the developers and users are expected to be performed with in the store,
some \appStores export part of the work to other platforms.
This can even occur for common features that one might find essential.
For instance, starred reviews are universal in Cluster 2, 4, and 6 where typical users leverage this information to decide whether an application is good; starred reviews are uncommon for other stores in Cluster 1, 5, 7.
The specialized store may have some other metric to indicate popularity or quality, such as total number of downloads, but the focus of the store is often to offer a managed way of installation.
Other features, such as application support, are left to other platforms such as social media.
Research can further explore the integration between \appStores and other platforms.

\section{Threats to Validity}
\label{sec:threats}

\paragraph{Internal Validity}
Our initial seeding of app stores comes from personal experience of app stores by the authors of the paper.
Personal bias could cause us to miss other types of app stores.
However, given the number of authors on this paper and our initial effort to consider as many stores as possible, we feel that have created a wide, deep, and collaborative ``best effort''.
When we labelled app stores by their dimensions, it is a qualitative process.
As with any qualitative process, the results could be biased by the authors performing the task.
We tackled this issue by first labeling a few stores separately by all authors and discussing the results until a consensus was achieved; thus, we started with a set of ``gold standard'' labels.
Then the labeling task was delegated to two authors who continued to label the stores separately with a portion of the store overlapping.
The overlapping labels are then verified by the \emph{Cohen's Kappa} between the two authors to measure the agreement.

We leveraged the \kmeans algorithm for the clustering process.
We first applied \emph{PCA} techniques to reduce the dimensions of the
initial labeling and provide an orthogonal basis to feed the \kmeans clustering.
When using other clustering algorithms (e.g., \emph{Mean-shift}, \emph{DBSCAN}), the clustering result might change;
while \kmeans is widely adopted for clustering process in SE research, by nature, determining the proper $k$ value is still a challenge.
We followed common best practice to use metrics (i.e., the \emph{Silhouette} method) to determine the best value $k$.
Despite our efforts, the output of the \kmeans clustering is not perfect.
We mainly leveraged the \kmeans clustering as the first step to illustrate that app stores forms natural clusters which are different from each other.
Base on the \kmeans output, we further grouped the clusters into types based on our qualitative understanding of the app store space.

\paragraph{External Validity}
During the process of expanding app stores, we relied on the \emph{Google Search Engine} to find web results based on keywords.
The results of this step rely on the capability of \emph{Google} and are subject to change over time as \emph{Google} updates its search algorithms.
The order may also be affected by SEO operations.
Combining results from other search engines (e.g., \emph{DuckDuckGo}, \emph{Bing}) can help to reduce the bias.

When we applied our inclusion criteria, 1) app stores must contain software products and 2) should offer an end-to-end experience for users (ordering, delivery, installation), we excluded stores that focus on digital assets that are not software, such as a pure assets store that offers cosmetic enhancements to desktop environments; we also excluded stores that offer software products but in a way such that installation is completely managed by users.
An extreme example, would be the software section of \emph{Amazon} where software is sold as an activation key which users would input to activate the software that they need to install themselves.
A more general inspection of all means of distribution software can be performed to gain a broader understanding of software distribution.

We relied on only publicly available information to label each store.
So if some functionality (e.g., analytics information) is not documented publicly, we were unable to confirm whether the store has such functionality.
We also set a time limit to label each store so in case we were unable to find information about the store, with each store receives the same amount of attention.

One of the main challenges for reproducibility and replicability is that the \emph{Google Search} results and app stores can change overtime. So in the future, if researchers would like to repeat our study, the labelling results may differ due to updates in the app store. To mitigate this issue, we've included a snapshot of all \emph{Google Search} results, and documented how we would perform the labelling. So while the final labels may differ, by applying the same process, a replication study would be possible with updated data.
 
\section{Summary}
\label{sec:summary}

In this paper, we have explored the idea of what an app store is and what features make app stores unique from each other.
We labelled a set of representative stores, curated from web search queries, by their features to study the natural groupings of the stores.
Our analysis suggests that app stores can differ in the type of product offered in the store, and whether the store is business oriented or community oriented.
These natural groupings of the stores challenge the manner in which app store research has largely been mobile focused.
Previous studies have already shown empirical differences in activities in mobile app stores and game stores~\cite{lin2019empirical}.
Our study further suggests that in the future, when we study app stores, we would need to consider the generalizability of the results across app stores.
Since one type of app store may operate under different constraints than another kind, results observed in one app store setting may not generalize to others.
 
\begin{acknowledgements}
  We would like to thank the attendees of the Shonan meeting~\cite{shonan.meeting.report} on ``Release Engineering for Mobile Applications'', where the paper's idea was conceived.

  One of the authors has received funding from the European Union's Horizon 2020 research and innovation programme under grant agreement number $825328$ (FASTEN).
\end{acknowledgements}

\bibliographystyle{ieeetr}
\bibliography{sample}

\begin{thebibliography}{100}

\bibitem{dixon2010home}
C.~Dixon, R.~Mahajan, S.~Agarwal, A.~Brush, B.~Lee, S.~Saroiu, and V.~Bahl,
  ``{The home needs an operating system (and an app store)},'' in {\em SIGCOMM
  Workshop on Hot Topics in Networks}, ACM, 2010.

\bibitem{store.steam}
Valve, ``{Welcome to Steam}.'' \url{https://store.steampowered.com/}, 2022.
\newblock Accessed: Jun. 22 2022.

\bibitem{store.github}
GitHub, ``{GitHub Marketplace · to improve your workflow · GitHub}.''
  \url{https://github.com/marketplace?type=}, 2022.
\newblock Accessed: Jun. 06 2022.

\bibitem{store.chrome}
Google, ``{Chrome Web Store - Extensions}.''
  \url{https://chrome.google.com/webstore/category/extensions}, 2022.
\newblock Accessed: Jun. 22, 2022.

\bibitem{store.wordpress}
WordPress, ``{WordPress Plugins | WordPress.org}.''
  \url{https://wordpress.org/plugins/}, 2022.
\newblock Accessed: Jun. 22, 2022.

\bibitem{store.autodesk}
Autodesk, ``{Autodesk App Store : Plugins, Add-ons for Autodesk software,
  AutoCAD, Revit, Inventor, 3ds Max, Maya ...}.''
  \url{https://apps.autodesk.com/}, 2022.
\newblock Accessed: Jun. 22, 2022.

\bibitem{store.dockerhub}
Docker, ``{Explore Docker's Container Image Repository | Docker Hub}.''
  \url{https://hub.docker.com/search?q=}, 2022.
\newblock Accessed: Jun. 22, 2022.

\bibitem{store.aws}
Amazon, ``{AWS Marketplace: Homepage}.''
  \url{https://aws.amazon.com/marketplace/}, 2022.
\newblock Accessed: Jun. 22, 2022.

\bibitem{store.homebrew}
{R\'emi Pr\'evost, Mike McQuaid, and Danielle Lalonde}, ``{The Missing Package
  Manager for macOS (or Linux) — Homebrew}.'' \url{https://brew.sh/}, 2022.
\newblock Accessed: Jun. 22, 2022.

\bibitem{store.ubuntu}
Canonical, ``{Ubuntu Software Center in Launchpad}.''
  \url{https://launchpad.net/software-center}, 2009.
\newblock Accessed: Jun. 22, 2022.

\bibitem{harman2012app}
M.~Harman, Y.~Jia, and Y.~Zhang, ``{App store mining and analysis: MSR for App
  Stores},'' in {\em Int. Conf. on Mining Software Repositories}, IEEE, 2012.

\bibitem{lin2019empirical}
D.~Lin, C.-P. Bezemer, Y.~Zou, and A.~E. Hassan, ``An empirical study of game
  reviews on the steam platform,'' in {\em Empirical Software Engineering},
  Springer, 2019.

\bibitem{wiki.appwrapper}
Wikipedia, ``{Electronic AppWrapper - Wikipedia}.''
  \url{https://en.wikipedia.org/wiki/Electronic_AppWrapper}, 2022.
\newblock Accessed: Jun. 22, 2022.

\bibitem{macqueen1967some}
J.~MacQueen {\em et~al.}, ``{Some methods for classification and analysis of
  multivariate observations},'' in {\em Proceedings of the fifth Berkeley
  symposium on mathematical statistics and probability}, Oakland, CA, USA,
  1967.

\bibitem{rousseeuw1987silhouettes}
P.~J. Rousseeuw, ``{Silhouettes: a graphical aid to the interpretation and
  validation of cluster analysis},'' in {\em Journal of computational and
  applied mathematics}, Elsevier, 1987.

\bibitem{ruiz2012understanding}
I.~J.~M. Ruiz, M.~Nagappan, B.~Adams, and A.~E. Hassan, ``{Understanding reuse
  in the android market},'' in {\em Int. Conf. on Program Comprehension}, IEEE,
  2012.

\bibitem{martin2016survey}
W.~Martin, F.~Sarro, Y.~Jia, Y.~Zhang, and M.~Harman, ``{A survey of app store
  analysis for software engineering},'' in {\em Transactions on Software
  Engineering}, IEEE, 2016.

\bibitem{dkabrowski2022analysing}
J.~D{\k{a}}browski, E.~Letier, A.~Perini, and A.~Susi, ``{Analysing app reviews
  for software engineering: a systematic literature review},'' in {\em
  Empirical Software Engineering}, Springer, 2022.

\bibitem{zhan2021atvhunter}
X.~Zhan, L.~Fan, S.~Chen, F.~Wu, T.~Liu, X.~Luo, and Y.~Liu, ``{Atvhunter:
  Reliable version detection of third-party libraries for vulnerability
  identification in android applications},'' in {\em Int. Conf. on Software
  Engineering}, IEEE, 2021.

\bibitem{zhang2020does}
X.~Zhang, X.~Wang, R.~Slavin, T.~Breaux, and J.~Niu, ``{How does
  misconfiguration of analytic services compromise mobile privacy?},'' in {\em
  Int. Conf. on Software Engineering}, IEEE, 2020.

\bibitem{rahaman2021algebraic}
S.~Rahaman, I.~Neamtiu, and X.~Yin, ``{Algebraic-datatype taint tracking, with
  applications to understanding Android identifier leaks},'' in {\em Joint
  Meeting on European Software Engineering Conference and Symposium on the
  Foundations of Software Engineering}, ACM, 2021.

\bibitem{nguyen2020code}
T.~Nguyen, P.~Vu, and T.~Nguyen, ``{Code recommendation for exception
  handling},'' in {\em Joint Meeting on European Software Engineering
  Conference and Symposium on the Foundations of Software Engineering}, ACM,
  2020.

\bibitem{pan2020static}
L.~Pan, B.~Cui, H.~Liu, J.~Yan, S.~Wang, J.~Yan, and J.~Zhang, ``{Static
  asynchronous component misuse detection for Android applications},'' in {\em
  Joint Meeting on European Software Engineering Conference and Symposium on
  the Foundations of Software Engineering}, ACM, 2020.

\bibitem{arzt2021sustainable}
S.~Arzt, ``{Sustainable Solving: Reducing The Memory Footprint of IFDS-Based
  Data Flow Analyses Using Intelligent Garbage Collection},'' in {\em Int.
  Conf. on Software Engineering}, IEEE, 2021.

\bibitem{yao2022describectx}
S.~Yang, Y.~Wang, Y.~Yao, H.~Wang, Y.~F. Ye, and X.~Xiao, ``{DescribeCtx:
  Context-Aware Description Synthesis for Sensitive Behaviors in Mobile
  Apps},'' in {\em Int. Conf. on Software Engineering}, IEEE, 2022.

\bibitem{dong2020time}
Z.~Dong, M.~B{\"o}hme, L.~Cojocaru, and A.~Roychoudhury, ``{Time-travel testing
  of android apps},'' in {\em Int. Conf. on Software Engineering}, IEEE, 2020.

\bibitem{chen2020empirical}
S.~Chen, L.~Fan, G.~Meng, T.~Su, M.~Xue, Y.~Xue, Y.~Liu, and L.~Xu, ``{An
  empirical assessment of security risks of global android banking apps},'' in
  {\em Int. Conf. on Software Engineering}, IEEE, 2020.

\bibitem{almanee2021too}
S.~Almanee, A.~{\"U}nal, M.~Payer, and J.~Garcia, ``{Too Quiet in the Library:
  An Empirical Study of Security Updates in Android Apps’ Native Code},'' in
  {\em Int. Conf. on Software Engineering}, IEEE, 2021.

\bibitem{alshayban2020accessibility}
A.~Alshayban, I.~Ahmed, and S.~Malek, ``{Accessibility issues in android apps:
  state of affairs, sentiments, and ways forward},'' in {\em Int. Conf. on
  Software Engineering}, IEEE, 2020.

\bibitem{yang2021don}
B.~Yang, Z.~Xing, X.~Xia, C.~Chen, D.~Ye, and S.~Li, ``{Don’t do that!
  hunting down visual design smells in complex uis against design
  guidelines},'' in {\em Int. Conf. on Software Engineering}, IEEE, 2021.

\bibitem{liu2021identifying}
P.~Liu, L.~Li, Y.~Yan, M.~Fazzini, and J.~Grundy, ``{Identifying and
  characterizing silently-evolved methods in the android API},'' in {\em Int.
  Conf. on Software Engineering: Software Engineering in Practice}, IEEE, 2021.

\bibitem{yu2021layout}
S.~Yu, C.~Fang, Y.~Yun, and Y.~Feng, ``{Layout and image recognition driving
  cross-platform automated mobile testing},'' in {\em Int. Conf. on Software
  Engineering}, IEEE, 2021.

\bibitem{ye2021empirical}
J.~Ye, K.~Chen, X.~Xie, L.~Ma, R.~Huang, Y.~Chen, Y.~Xue, and J.~Zhao, ``{An
  empirical study of GUI widget detection for industrial mobile games},'' in
  {\em Joint Meeting on European Software Engineering Conference and Symposium
  on the Foundations of Software Engineering}, ACM, 2021.

\bibitem{ma2021fine}
S.~Ma, J.~Li, H.~Kim, E.~Bertino, S.~Nepal, D.~Ostry, and C.~Sun, ``{Fine with
  “1234”? An Analysis of SMS One-Time Password Randomness in Android
  Apps},'' in {\em Int. Conf. on Software Engineering}, IEEE, 2021.

\bibitem{song2021imgdroid}
W.~Song, M.~Han, and J.~Huang, ``{IMGDroid: Detecting Image Loading Defects in
  Android Applications},'' in {\em Int. Conf. on Software Engineering}, IEEE,
  2021.

\bibitem{zhao2021guigan}
T.~Zhao, C.~Chen, Y.~Liu, and X.~Zhu, ``{GUIGAN: Learning to Generate GUI
  Designs Using Generative Adversarial Networks},'' in {\em Int. Conf. on
  Software Engineering}, IEEE, 2021.

\bibitem{chen2020unblind}
J.~Chen, C.~Chen, Z.~Xing, X.~Xu, L.~Zhut, G.~Li, and J.~Wang, ``{Unblind your
  apps: Predicting natural-language labels for mobile gui components by deep
  learning},'' in {\em Int. Conf. on Software Engineering}, IEEE, 2020.

\bibitem{kuznetsov2021frontmatter}
K.~Kuznetsov, C.~Fu, S.~Gao, D.~N. Jansen, L.~Zhang, and A.~Zeller,
  ``{Frontmatter: mining Android user interfaces at scale},'' in {\em Joint
  Meeting on European Software Engineering Conference and Symposium on the
  Foundations of Software Engineering}, ACM, 2021.

\bibitem{van2020schrodinger}
D.~Van Der~Linden, P.~Anthonysamy, B.~Nuseibeh, T.~T. Tun, M.~Petre, M.~Levine,
  J.~Towse, and A.~Rashid, ``{Schr{\"o}dinger's security: Opening the box on
  app developers' security rationale},'' in {\em Int. Conf. on Software
  Engineering}, IEEE, 2020.

\bibitem{murali2021scalable}
V.~Murali, E.~Yao, U.~Mathur, and S.~Chandra, ``{Scalable statistical root
  cause analysis on app telemetry},'' in {\em Int. Conf. on Software
  Engineering: Software Engineering in Practice}, IEEE, 2021.

\bibitem{sun2021empirical}
R.~Sun, W.~Wang, M.~Xue, G.~Tyson, S.~Camtepe, and D.~C. Ranasinghe, ``{An
  empirical assessment of global COVID-19 contact tracing applications},'' in
  {\em Int. Conf. on Software Engineering}, IEEE, 2021.

\bibitem{truelove2021we}
A.~Truelove, E.~S. de~Almeida, and I.~Ahmed, ``{We’ll Fix It in Post: What Do
  Bug Fixes in Video Game Update Notes Tell Us?},'' in {\em Int. Conf. on
  Software Engineering}, IEEE, 2021.

\bibitem{haering2021automatically}
M.~Haering, C.~Stanik, and W.~Maalej, ``{Automatically matching bug reports
  with related app reviews},'' in {\em Int. Conf. on Software Engineering},
  IEEE, 2021.

\bibitem{yu2021prioritize}
S.~Yu, C.~Fang, Z.~Cao, X.~Wang, T.~Li, and Z.~Chen, ``{Prioritize crowdsourced
  test reports via deep screenshot understanding},'' in {\em Int. Conf. on
  Software Engineering}, IEEE, 2021.

\bibitem{obie2021first}
H.~O. Obie, W.~Hussain, X.~Xia, J.~Grundy, L.~Li, B.~Turhan, J.~Whittle, and
  M.~Shahin, ``{A first look at human values-violation in app reviews},'' in
  {\em Int. Conf. on Software Engineering: Software Engineering in Society},
  IEEE, 2021.

\bibitem{fischer2021does}
R.~A.-L. Fischer, R.~Walczuch, and E.~Guzman, ``{Does culture matter? impact of
  individualism and uncertainty avoidance on app reviews},'' in {\em Int. Conf.
  on Software Engineering: Software Engineering in Society}, IEEE, 2021.

\bibitem{haggag2021covid}
O.~Haggag, S.~Haggag, J.~Grundy, and M.~Abdelrazek, ``{COVID-19 vs social media
  apps: does privacy really matter?},'' in {\em Int. Conf. on Software
  Engineering: Software Engineering in Society}, IEEE, 2021.

\bibitem{shams2020society}
R.~A. Shams, W.~Hussain, G.~Oliver, A.~Nurwidyantoro, H.~Perera, and
  J.~Whittle, ``{Society-oriented applications development: Investigating
  users’ values from bangladeshi agriculture mobile applications},'' in {\em
  Int. Conf. on Software Engineering: Software Engineering in Society}, IEEE,
  2020.

\bibitem{zhang2021checking}
Z.~Zhang, Y.~Feng, M.~D. Ernst, S.~Porst, and I.~Dillig, ``{Checking
  conformance of applications against GUI policies},'' in {\em Joint Meeting on
  European Software Engineering Conference and Symposium on the Foundations of
  Software Engineering}, ACM, 2021.

\bibitem{wu2021identifying}
H.~Wu, W.~Deng, X.~Niu, and C.~Nie, ``{Identifying key features from app user
  reviews},'' in {\em Int. Conf. on Software Engineering}, IEEE, 2021.

\bibitem{hu2021champ}
Y.~Hu, H.~Wang, T.~Ji, X.~Xiao, X.~Luo, P.~Gao, and Y.~Guo, ``{Champ:
  Characterizing undesired app behaviors from user comments based on market
  policies},'' in {\em Int. Conf. on Software Engineering}, IEEE, 2021.

\bibitem{guo2020caspar}
H.~Guo and M.~P. Singh, ``{Caspar: extracting and synthesizing user stories of
  problems from app reviews},'' in {\em Int. Conf. on Software Engineering},
  IEEE, 2020.

\bibitem{al-subaihin2021app}
A.~A. {Al-Subaihin}, F.~Sarro, S.~Black, L.~Capra, and M.~Harman, ``{App Store
  Effects on Software Engineering Practices},'' in {\em Transactions on
  Software Engineering}, IEEE, 2021.

\bibitem{wang2018beyond}
H.~Wang, Z.~Liu, J.~Liang, N.~Vallina-Rodriguez, Y.~Guo, L.~Li, J.~Tapiador,
  J.~Cao, and G.~Xu, ``{Beyond google play: A large-scale comparative study of
  chinese android app markets},'' in {\em Internet Measurement Conference
  2018}, 2018.

\bibitem{jansen2013defining}
S.~Jansen and E.~Bloemendal, ``{Defining app stores: The role of curated
  marketplaces in software ecosystems},'' in {\em ICSOB}, Springer, 2013.

\bibitem{walker2006grounded}
D.~Walker and F.~Myrick, ``{Grounded theory: An exploration of process and
  procedure},'' in {\em Qualitative health research}, Sage, 2006.

\bibitem{coxon1999sorting}
A.~P.~M. Coxon {\em et~al.}, {\em Sorting data: Collection and analysis}.
\newblock Sage, 1999.

\bibitem{adolph2011using}
S.~Adolph, W.~Hall, and P.~Kruchten, ``{Using grounded theory to study the
  experience of software development},'' in {\em Empirical Software
  Engineering}, Springer, 2011.

\bibitem{hoda2012developing}
R.~Hoda, J.~Noble, and S.~Marshall, ``{Developing a grounded theory to explain
  the practices of self-organizing Agile teams},'' in {\em Empirical Software
  Engineering}, Springer, 2012.

\bibitem{masood2020agile}
Z.~Masood, R.~Hoda, and K.~Blincoe, ``{How agile teams make self-assignment
  work: a grounded theory study},'' in {\em Empirical Software Engineering},
  Springer, 2020.

\bibitem{vassallo2020developers}
C.~Vassallo, S.~Panichella, F.~Palomba, S.~Proksch, H.~C. Gall, and A.~Zaidman,
  ``{How developers engage with static analysis tools in different contexts},''
  in {\em Empirical Software Engineering}, Springer, 2020.

\bibitem{chen2021maintenance}
J.~Chen, X.~Xia, D.~Lo, J.~Grundy, and X.~Yang, ``{Maintenance-related concerns
  for post-deployed Ethereum smart contract development: issues, techniques,
  and future challenges},'' in {\em Empirical Software Engineering}, Springer,
  2021.

\bibitem{wang2022demystifying}
P.~Wang, C.~Brown, J.~A. Jennings, and K.~T. Stolee, ``{Demystifying regular
  expression bugs},'' in {\em Empirical Software Engineering}, Springer, 2022.

\bibitem{cohen1960coefficient}
J.~Cohen, ``{A coefficient of agreement for nominal scales},'' in {\em
  Educational and psychological measurement}, Sage, 1960.

\bibitem{perez2020systematic}
J.~P{\'e}rez, J.~D{\'\i}az, J.~Garcia-Martin, and B.~Tabuenca, ``{Systematic
  literature reviews in software engineering—Enhancement of the study
  selection process using Cohen’s kappa statistic},'' in {\em Journal of
  Systems and Software}, Elsevier, 2020.

\bibitem{lantz1996behavior}
C.~A. Lantz and E.~Nebenzahl, ``{Behavior and interpretation of the $\kappa$
  statistic: Resolution of the two paradoxes},'' in {\em Journal of clinical
  epidemiology}, Elsevier, 1996.

\bibitem{pickerill2020phantom}
P.~Pickerill, H.~J. Jungen, M.~Ochodek, M.~Ma{\'c}kowiak, and M.~Staron,
  ``Phantom: Curating github for engineered software projects using time-series
  clustering,'' {\em Empirical Software Engineering}, 2020.

\bibitem{khatibi2014flexible}
V.~Khatibi~Bardsiri, D.~N.~A. Jawawi, S.~Z.~M. Hashim, and E.~Khatibi, ``A
  flexible method to estimate the software development effort based on the
  classification of projects and localization of comparisons,'' {\em Empirical
  Software Engineering}, 2014.

\bibitem{al2019empirical}
A.~Al-Subaihin, F.~Sarro, S.~Black, and L.~Capra, ``Empirical comparison of
  text-based mobile apps similarity measurement techniques,'' {\em Empirical
  Software Engineering}, 2019.

\bibitem{kuchta2018correctness}
T.~Kuchta, T.~Lutellier, E.~Wong, L.~Tan, and C.~Cadar, ``{On the correctness
  of electronic documents: studying, finding, and localizing inconsistency bugs
  in PDF readers and files},'' {\em Empirical Software Engineering}, 2018.

\bibitem{arthur2006k}
D.~Arthur and S.~Vassilvitskii, ``k-means++: The advantages of careful
  seeding,'' tech. rep., Stanford, 2006.

\bibitem{wold1987principal}
S.~Wold, K.~Esbensen, and P.~Geladi, ``{Principal component analysis},'' in
  {\em Chemometrics and intelligent laboratory systems}, Elsevier, 1987.

\bibitem{app.store.introduce}
Apple, ``{Apple Introduces the New iPhone 3G}.''
  \url{https://www.apple.com/ca/newsroom/2008/06/09Apple-Introduces-the-New-iPhone-3G/},
  2008.
\newblock Accessed: Jul. 17, 2022.

\bibitem{chrome.disable.payment}
Google, ``{Chrome Web Store payments deprecation}.''
  \url{https://developer.chrome.com/docs/webstore/cws-payments-deprecation/},
  2022.
\newblock Accessed: Mar. 16, 2022.

\bibitem{eu.commission}
E.~Commission, ``{Digital Markets Act: Commission welcomes political agreement
  on rules to ensure fair and open digital markets}.''
  \url{https://ec.europa.eu/commission/presscorner/detail/en/IP_22_1978}, 2022.
\newblock Accessed: Jul. 13, 2022.

\bibitem{store.fdroid}
F-Droid, ``{F-Droid - Free and Open Source Android App Repository}.''
  \url{https://f-droid.org/}, 2022.
\newblock Accessed: Oct. 02, 2022.

\bibitem{org.npm}
npm, ``{npm About}.'' \url{https://www.npmjs.com/about}, 2022.
\newblock Accessed: Oct. 02, 2022.

\bibitem{org.javascript}
E.~International, ``{TC39 - Specifying JavaScript.}.'' \url{https://tc39.es/},
  2022.
\newblock Accessed: Oct. 02, 2022.

\bibitem{kodi.tmdb}
T.~Kodi, ``{The Movie Database Python | Matrix | Addons | Kodi}.''
  \url{https://kodi.tv/addons/matrix/metadata.themoviedb.org.python}, 2022.
\newblock Accessed: Jul. 13, 2022.

\bibitem{pham2017paas}
V.~V.~H. Pham, X.~Liu, X.~Zheng, M.~Fu, S.~V. Deshpande, W.~Xia, R.~Zhou, and
  M.~Abdelrazek, ``{PaaS-black or white: an investigation into software
  development model for building retail industry SaaS},'' in {\em Int. Conf. on
  Software Engineering Companion (ICSE-C)}, IEEE, 2017.

\bibitem{finkelstein2017investigating}
A.~Finkelstein, M.~Harman, Y.~Jia, W.~Martin, F.~Sarro, and Y.~Zhang,
  ``{Investigating the relationship between price, rating, and popularity in
  the Blackberry World App Store},'' {\em Information and Software Technology},
  2017.

\bibitem{sarro2015feature}
F.~Sarro, A.~A. Al-Subaihin, M.~Harman, Y.~Jia, W.~Martin, and Y.~Zhang,
  ``{Feature lifecycles as they spread, migrate, remain, and die in app
  stores},'' in {\em Int. requirements engineering conference (RE)}, IEEE,
  2015.

\bibitem{aljedaani2019comparison}
W.~Aljedaani, M.~Nagappan, B.~Adams, and M.~Godfrey, ``{A comparison of bugs
  across the ios and android platforms of two open source cross platform
  browser apps},'' in {\em Int. Conf. on Mobile Software Engineering and
  Systems}, IEEE, 2019.

\bibitem{lu2019blockchain}
Z.~Lu, Y.~Shi, R.~Tao, and Z.~Zhang, ``Blockchain for digital rights management
  of design works,'' in {\em Int. Conf on Software Engineering and Service
  Science (ICSESS)}, IEEE, 2019.

\bibitem{gaber2020privdrm}
T.~Gaber, A.~Ahmed, and A.~Mostafa, ``Privdrm: A privacy-preserving secure
  digital right management system,'' in {\em Evaluation and Assessment in
  Software Engineering}, ACM, 2020.

\bibitem{sung2019test}
A.~Sung, S.~Kim, Y.~Kim, Y.~Jang, and J.~Kim, ``Test automation and its
  limitations: a case study,'' in {\em Int. Conf. on Automated Software
  Engineering (ASE)}, IEEE, 2019.

\bibitem{denuvo.performance}
M.~Lemon, ``{Two Point Hospital no longer uses Denuvo DRM}.''
  \url{https://www.vg247.com/two-point-hospital-no-longer-uses-denuvo-drm},
  2018.
\newblock Accessed: Mar. 31, 2023.

\bibitem{he2015mining}
X.~He, W.~Dai, G.~Cao, R.~Tang, M.~Yuan, and Q.~Yang, ``{Mining target users
  for online marketing based on app store data},'' in {\em Int. Conf. on Big
  Data (Big Data)}, IEEE, 2015.

\bibitem{scoccia2022empirical}
G.~L. Scoccia, M.~Autili, G.~Stilo, and P.~Inverardi, ``{An empirical study of
  privacy labels on the Apple iOS mobile app store},'' in {\em Int. Conf. on
  Mobile Software Engineering and Systems}, 2022.

\bibitem{murphy2014cowboys}
E.~Murphy-Hill, T.~Zimmermann, and N.~Nagappan, ``{Cowboys, ankle sprains, and
  keepers of quality: How is video game development different from software
  development?},'' in {\em Int. Conf. on Software Engineering}, 2014.

\bibitem{pascarella2018video}
L.~Pascarella, F.~Palomba, M.~Di~Penta, and A.~Bacchelli, ``{How is video game
  development different from software development in open source?},'' in {\em
  Int. Conf. on Mining Software Repositories}, IEEE, 2018.

\bibitem{lee2020empirical}
D.~Lee, G.~K. Rajbahadur, D.~Lin, M.~Sayagh, C.-P. Bezemer, and A.~E. Hassan,
  ``{An empirical study of the characteristics of popular Minecraft mods},''
  {\em Empirical Software Engineering}, 2020.

\bibitem{ibrahim2020too}
M.~H. Ibrahim, M.~Sayagh, and A.~E. Hassan, ``{Too many images on dockerhub!
  how different are images for the same system?},'' {\em Empirical Software
  Engineering}, 2020.

\bibitem{subramanian2006empirical}
G.~H. Subramanian, P.~C. Pendharkar, and M.~Wallace, ``{An empirical study of
  the effect of complexity, platform, and program type on software development
  effort of business applications},'' {\em Empirical Software Engineering},
  2006.

\bibitem{manotas2016empirical}
I.~Manotas, C.~Bird, R.~Zhang, D.~Shepherd, C.~Jaspan, C.~Sadowski, L.~Pollock,
  and J.~Clause, ``{An empirical study of practitioners' perspectives on green
  software engineering},'' in {\em Int. Conf. on Software Engineering}, 2016.

\bibitem{gholami2021should}
S.~Gholami, H.~Khazaei, and C.-P. Bezemer, ``{Should you upgrade official
  docker hub images in production environments?},'' in {\em Int. Conf. on
  Software Engineering: New Ideas and Emerging Results (ICSE-NIER)}, IEEE,
  2021.

\bibitem{guzman2018user}
E.~Guzman, L.~Oliveira, Y.~Steiner, L.~C. Wagner, and M.~Glinz, ``{User
  feedback in the app store: a cross-cultural study},'' in {\em Int. Conf. on
  Software Engineering: Software Engineering in Society}, 2018.

\bibitem{lee2020building}
D.~Lee, D.~Lin, C.-P. Bezemer, and A.~E. Hassan, ``{Building the perfect
  game--an empirical study of game modifications},'' {\em Empirical Software
  Engineering}, 2020.

\bibitem{nayebi2017version}
M.~Nayebi, H.~Farahi, and G.~Ruhe, ``{Which version should be released to app
  store?},'' in {\em Int. Symposium on Empirical Software Engineering and
  Measurement (ESEM)}, IEEE, 2017.

\bibitem{nayebi2016release}
M.~Nayebi, B.~Adams, and G.~Ruhe, ``{Release Practices for Mobile Apps--What do
  Users and Developers Think?},'' in {\em Int. Conf. on software analysis,
  evolution, and reengineering (saner)}, IEEE, 2016.

\bibitem{shen2017towards}
S.~Shen, X.~Lu, Z.~Hu, and X.~Liu, ``{Towards release strategy optimization for
  apps in Google Play},'' in {\em Proceedings of the 9th Asia-Pacific Symposium
  on Internetware}, 2017.

\bibitem{ferreira2021containing}
G.~Ferreira, L.~Jia, J.~Sunshine, and C.~K{\"a}stner, ``{Containing malicious
  package updates in npm with a lightweight permission system},'' in {\em Int.
  Conf. on Software Engineering (ICSE)}, IEEE, 2021.

\bibitem{tang2019large}
C.~Tang, S.~Chen, L.~Fan, L.~Xu, Y.~Liu, Z.~Tang, and L.~Dou, ``{A large-scale
  empirical study on industrial fake apps},'' in {\em Int. Conf. on Software
  Engineering: Software Engineering in Practice (ICSE-SEIP)}, IEEE, 2019.

\bibitem{vouillon2013software}
J.~Vouillon and R.~D. Cosmo, ``{On software component co-installability},''
  {\em Transactions on Software Engineering and Methodology (TOSEM)}, 2013.

\bibitem{claes2015historical}
M.~Claes, T.~Mens, R.~Di~Cosmo, and J.~Vouillon, ``{A historical analysis of
  Debian package incompatibilities},'' in {\em Int. Conf. on Mining Software
  Repositories}, IEEE, 2015.

\bibitem{mcmillan2012detecting}
C.~McMillan, M.~Grechanik, and D.~Poshyvanyk, ``{Detecting similar software
  applications},'' in {\em Int. Conf. on Software Engineering (ICSE)}, IEEE,
  2012.

\bibitem{martin2016causal}
W.~Martin, F.~Sarro, and M.~Harman, ``{Causal impact analysis for app releases
  in google play},'' in {\em Int. Symposium on Foundations of software
  engineering}, 2016.

\bibitem{maalej2019data}
W.~Maalej, M.~Nayebi, and G.~Ruhe, ``{Data-driven requirements engineering-an
  update},'' in {\em Int. Conf. on Software Engineering: Software Engineering
  in Practice (ICSE-SEIP)}, IEEE, 2019.

\bibitem{ullmann2022makes}
G.~C. Ullmann, C.~Politowski, Y.-G. Gu{\'e}h{\'e}neuc, and F.~Petrillo, ``{What
  makes a game high-rated? towards factors of video game success},'' in {\em
  Int. ICSE Workshop on Games and Software Engineering: Engineering Fun,
  Inspiration, and Motivation}, 2022.

\bibitem{khalid2014prioritizing}
H.~Khalid, M.~Nagappan, E.~Shihab, and A.~E. Hassan, ``{Prioritizing the
  devices to test your app on: A case study of android game apps},'' in {\em
  Int. Symposium on Foundations of Software Engineering}, 2014.

\bibitem{nayebi2017app}
M.~Nayebi, H.~Cho, H.~Farrahi, and G.~Ruhe, ``{App store mining is not
  enough},'' in {\em Int. Conf. on Software Engineering Companion (ICSE-C)},
  IEEE, 2017.

\bibitem{rosen2016mobile}
C.~Rosen and E.~Shihab, ``{What are mobile developers asking about? a large
  scale study using stack overflow},'' in {\em Empirical Software Engineering},
  Springer, 2016.

\bibitem{wang2019characterizing}
H.~Wang, X.~Wang, and Y.~Guo, ``{Characterizing the global mobile app
  developers: a large-scale empirical study},'' in {\em Int. Conf. on Mobile
  Software Engineering and Systems}, IEEE, 2019.

\bibitem{ng2014android}
Y.~Y. Ng, H.~Zhou, Z.~Ji, H.~Luo, and Y.~Dong, ``{Which Android app store can
  be trusted in China?},'' in {\em Computer Software and Applications
  Conference}, IEEE, 2014.

\bibitem{shonan.meeting.report}
S.~McIntosh, Y.~Kamei, and M.~Nagappan, {\em {Release Engineering for Mobile
  Applications --- Communications of NII Shonan Meetings}}.
\newblock Springer, 2019.

\end{thebibliography}

\end{document}